\documentclass[]{article}

\usepackage{times}
\usepackage{pgfplots,tikz-3dplot}
\usepackage{amsmath}
\usepackage{multirow}
\usepackage{hyperref}
\usepackage{ar}

\newcommand{\aver}[1]{ \! \left\langle {#1} \right \rangle \!}
\newcommand{\vect}[1]{\mathbf{#1}}

\begin{document}

\title{The importance of corner sharpness \\ in the BARC test case: \\ a numerical study}
\author{Alessandro Chiarini, Maurizio Quadrio \\
        Politecnico di Milano, Dept. Aerospace Science and Technologies
}

\date{\today}

\maketitle

\begin{abstract}
The BARC flow is studied via Direct Numerical Simulation at a relatively low turbulent Reynolds number, with focus on the geometrical representation of the leading-edge (LE) corners. The study contributes to further our understanding of the discrepancies between existing numerical and experimental BARC data. In a first part, rounded LE corners with small curvature radii are considered. Results show that a small amount of rounding does not lead to abrupt changes of the mean fields, but that the effects increase with the curvature radius. The shear layer separates from the rounded LE at a lower angle, which reduces the size of the main recirculating region over the cylinder side. In contrast, the longitudinal size of the recirculating region behind the trailing edge (TE) increases, as the TE shear layer is accelerated. The effect of the curvature radii on the turbulent kinetic energy and on its production, dissipation and transport are addressed. The present results should be contrasted with the recent work of \cite{rocchio-etal-2020}, who found via implicit Large-Eddy Simulations at larger Reynolds numbers than even a small curvature radius leads to significant changes of the mean flow.

In a second part, the LE corners are fully sharp and the exact analytical solution of the Stokes problem in the neighbourhood of the corner is used to locally restore the solution accuracy degraded by the singularity. Changes in the mean flow reveal that the analytical correction leads to streamlines that better follow the corners. The flow separates from the LE with a lower angle, resulting in a slightly smaller recirculating region. The corner-correction approach is valuable in general, and is expected to help developing high-quality numerical simulations at the high Reynolds numbers typical of the experiments with reasonable meshing requirements.
\end{abstract}

\section{Introduction}

The flow around bluff bodies with sharp corners is interesting for both fundamental research and industrial applications, particularly in civil engineering. The rectangular cylinder is a simple yet representative prototype of such bodies. Despite the simple geometry, the flow around a rectangular cylinder contains a rich physics with several coexisting phenomena: a corner-induced separation, a detached boundary layer that may become unstable and reattach downstream, several recirculating regions and a large wake. Depending on the aspect ratio $\AR=L/D$ (where $L$ and $D$ are the longitudinal and vertical sizes of the body), the rectangular cylinder spans the overall range of blunt bodies from a flat plate normal to the flow ($\AR \rightarrow 0$), to a square cylinder ($\AR = 1$) and to a flat plate parallel to the flow ($\AR \rightarrow \infty$). At low values of the Reynolds number $Re$ and for $\AR \ge 3$, leading-edge (LE) and trailing-edge (TE) vortex shedding are interlocked to a unique frequency as a result of the interaction between the impinging shear layer instability and the TE shedding \cite{hourigan-etal-1993,hourigan-thompson-tan-2001,mills-etal-2002,mills-etal-2003}. When $\AR$ is increased, the Strouhal number based on $L$ and the incoming velocity faces an (almost) stepwise increase, depending on the number of vortices shed by the LE shear layer that are simultaneously present over the cylinder side; see \cite{okajima-1982,nakamura-nakashima-1986,ozono-etal-1992,tan-thompson-hourigan-1998}. At higher $Re$, the picture complicates even further, with large-scale vortices coexisting and interacting with small-scale turbulent fluctuations.

The value $\AR=5$ defines the international benchmark known as BARC (Benchmark of the Aerodynamics of a Rectangular 5:1 Cylinder) \cite{bartoli-etal-2008}. The goal of the BARC is to develop best practices for both experiments and simulations and to qualitatively and quantitatively characterise the main features of the flow, such as the shedding frequency and the main recirculating regions of the mean flow. An overview of the contributions to the BARC benchmark is provided in \cite{bruno-salvetti-ricciardelli-2014} and in the more recent works by \cite{patruno-etal-2016,mariotti-etal-2016,mariotti-siconolfi-salvetti-2017,mannini-etal-2017,ricci-etal-2017,moore-etal-2019}. Bruno et al. in \cite{bruno-salvetti-ricciardelli-2014} pointed out that, despite the fixed value of $\AR$, a significant variability of the available data remains, that complicates the flow characterisation. This is due to the strong sensitivity of the flow to several aspects of both experiments and numerical simulations (mainly RANS and LES studies). For the experiments, critical issues are e.g. measurements uncertainties, geometrical imperfections of the model, free stream turbulence. For numerical simulations, instead, we mention RANS and LES turbulence modelling, various  discretisation choices and the numerical method itself. In a scenario of highly scattered data, \cite{cimarelli-leonforte-angeli-2018b} performed the first Direct Numerical Simulation (DNS) of the BARC flow in turbulent regime employing the finite-volume toolbox OpenFOAM \cite{weller-etal-1998}, at relatively low Reynolds number $Re=3000$. Our own contribution \cite{chiarini-quadrio-2021} consisted in replicating that work by using an in-house finite-difference code and a finer grid to assess the robustness of the available results.

BARC data from numerical simulations tend to deviate significantly from those obtained with experimental measurements. For example, \cite{bruno-salvetti-ricciardelli-2014} report that the recirculation region over the cylinder side differs, with highly-resolved LES simulations not matching experiments with their prediction of a consistently shorter recirculation. Besides turbulence modelling, numerical discretisation and experimental uncertainties, such discrepancies may derive from differences in the setup, i.e. the boundary conditions at the inlet and spanwise boundaries and the inherent geometrical imperfections of the experimental models. In the numerical simulations the inlet condition is that of an unperturbed flow parallel to the rectangular cylinder without free-stream turbulence, but in experiments the incoming flow is less deterministic than that. This aspect was already found by Mannini et al. \cite{mannini-etal-2017} to affect the longitudinal extent of the main recirculating region. They studied experimentally the effect of free-stream turbulence and of the angle of attack on the main features of the BARC flow, finding that an increase of the free-stream turbulence shifts upstream the peak of the root-mean-square value of the pressure coefficient, implying a decrease of the length of the main recirculating region over the cylinder \cite{kiya-sasaki-1983}. In the spanwise direction, almost every numerical simulation employs periodic boundary conditions, which are clearly different from a wind-tunnel experiment where a solid wall exists. However, Bruno et al. \cite{bruno-coste-fransos-2012} have shown that this boundary condition does not significantly affect the size of the main recirculating bubble. 

The present work focuses on the geometrical representation of the corners. Perfectly sharp corners are obviously just an idealisation: a laboratory model will always have corners affected to some extent by manufacturing inaccuracies. The effect of rounding the corners has been mostly studied for a square cylinder (where the recirculation region on the side is missing) by for example \cite{park-yang-2016} and \cite{cao-tamura-2017} for low and high Reynolds numbers. For rectangular cylinders with larger $\AR$ we recall the works \cite{lamballais-etal-2008,lamballais-etal-2010} that investigate both three-dimensional and two-dimensional infinite D-shaped bodies changing the curvature radius $R$ of the upstream corners. \cite{cimarelli-franciolini-crivellini-2020} investigated the effects of different geometrical peculiarities such as the LE corners and the presence of a TE on the main flow features. 
\cite{chiarini-etal-2020} studied how rounded LE and TE corners affect the occurrence of the first instability. 
Recently, \cite{rocchio-etal-2020} used an implicit LES to carry out a sensitivity analysis of the BARC flow to the rounding of the LE corners. They identified the inadequate treatment of the corners as one of the reasons for the disagreement between numerical and experimental data. Indeed, they found that introducing even a tiny curvature radius is enough to significantly enlarge the size of the main recirculating region, thus reducing the discrepancy with the experimental data. However, the generality of this result with respect to both the Reynolds number and the numerical approach remains to be ascertained, and this is one of the two goals of the present contribution.

The other goal is addressing the accuracy losses induced in a numerical simulation of the BARC flow by the presence of a perfectly sharp corner. An ideal geometry can be easily achieved in a numerical simulation, but the geometrical singularity of the corner leads to a mathematical singularity for the Navier--Stokes equations, with extremely large velocity derivatives. As a result, the computational grid must be extremely fine near the corner to avoid a local accuracy drop. As an example, in \cite{chiarini-quadrio-2021} the DNS by \cite{cimarelli-leonforte-angeli-2018b} was repeated with a finer grid and the fluctuations of the lift coefficient were found to be much larger, because the finer grid captures better the flow separation. This is also supported by the laminar results of \cite{sohankar-norberg-davidson-1998} and \cite{anzai-etal-2017}. Unfortunately, the requirement of a very fine grid becomes more difficult to satisfy as $Re$ is increased, and renders DNS simulations at the large Reynolds numbers typical of the experiments extremely expensive if not impossible. There are however workarounds to cope with the corner singularity, so that the local solution accuracy can be maintained without resorting to excessively fine grids; see for example \cite{auteri-parolini-quartapelle-2003}. Among them we recall those based on the concept, first described by Moffat in \cite{moffat-1964}, that in the vicinity of the corners the flow is well described by the Stokes equations, which can be dealt with analytically. Previous examples where the Stokes solution was successfully used to improve the Navier--Stokes solution near corners include classic fluid dynamics problems such as the square cavity, see for example \cite{luchini-1991}.

This work consists in a DNS study of the turbulent BARC flow where the focus is on the LE corners. In a first part, the work \cite{rocchio-etal-2020} is reconsidered with DNS, and the effect of rounded LE corners is studied by considering two small curvature radii, i.e. $R/D=1/128$ and $R/D=1/64$. Compared to the reference work, we use a higher-fidelity numerical approach without modelling error, but at the cost of a lower Reynolds number, which is set to $Re=3000$ as in our previous DNS \cite{chiarini-quadrio-2021} used here as baseline. In a second part, the corners are considered to be sharp, but we employ the method described in \cite{luchini-1991} for corner correction, applied to the BARC flow for the first time. The method is presented in a form tailored to the BARC geometry, and the changes on the main features of the flow are discussed.

\section{The computational approach}
\label{sec:methods}

\begin{figure}
\centering
	\tdplotsetmaincoords{60}{135}	
	\begin{tikzpicture}[scale=0.30,tdplot_main_coords]
	
	\tikzset{myptr/.style={decoration={markings,mark=at position 1 with {\arrow[scale=3,>=stealth]{>}}},postaction={decorate}}}

        \coordinate (Aup) at ( -6, -8,  0.5);
        \coordinate (Bup) at ( 2, -8,  0.5);
        \coordinate (Cup) at ( 2,  8,  0.5);
        \coordinate (Dup) at ( -6,  8,  0.5);
        \coordinate (Alo) at ( -6, -8, -0.5);
        \coordinate (Blo) at ( 2, -8, -0.5);
        \coordinate (Clo) at ( 2,  8, -0.5);
        \coordinate (Dlo) at ( -6,  8, -0.5);

        \coordinate (A1) at ( 2,  9.5, -0.5);
        \coordinate (A2) at (-6,  9.5, -0.5);
        \coordinate (A3) at ( 2,    8, -0.5);
        \coordinate (A4) at (-6,    8, -0.5);
        \draw[<->] (A1)--(A2);
        \draw[] (A1)--(A3);
        \draw[] (A2)--(A4);
        \node at (-0.5,11,-0.5) {$L=5D$};
        
        \draw[dotted] (2,8,-0,5) -- (2,8,-6);
        
        \coordinate (B1) at ( -6.5, 8, 0.5);
        \coordinate (B2) at ( -6.5, 8, -0.5);
        \coordinate (B3) at ( -6,   8, 0.5);
        \coordinate (B4) at ( -6,   8, -0.5);
        \draw[<->] (B1)--(B2);
        \draw[] (B1)--(B3);
        \draw[] (B2)--(B4);
        \node at (-7.7,8,0) {$D$};

        \draw[very thick,black] (Aup) -- (Bup);
        \draw[very thick,black] (Bup) -- (Cup);
        \draw[very thick,black] (Cup) -- (Dup);
        \draw[very thick,black] (Dup) -- (Aup);

        \draw[very thick,black,dashed] (Alo) -- (Blo);
        \draw[very thick,black] (Blo) -- (Clo);
        \draw[very thick,black] (Clo) -- (Dlo);
        \draw[very thick,black,dashed] (Dlo) -- (Alo);

        \draw[very thick,black,dashed] (Aup) -- (Alo);
        \draw[very thick,black] (Bup) -- (Blo);
        \draw[very thick,black] (Cup) -- (Clo);
        \draw[very thick,black] (Dup) -- (Dlo);

        \coordinate (AAup) at (-19, -8,  0.5);
        \coordinate (BBup) at ( 12, -8,  0.5);
        \coordinate (CCup) at ( 12,  8,  0.5);
        \coordinate (DDup) at (-19,  8,  0.5);
        \coordinate (AAlo) at (-19, -8, -0.5);
        \coordinate (BBlo) at ( 12, -8, -0.5);
        \coordinate (CClo) at ( 12,  8, -0.5);
        \coordinate (DDlo) at (-19,  8, -0.5);

        \coordinate (AAupp) at (  -6, -8,  6);
        \coordinate (BBupp) at (  2, -8,  6);
        \coordinate (CCupp) at (  2,  8,  6);
        \coordinate (DDupp) at (  -6,  8,  6);
        \coordinate (AAloo) at (  2, -8, -6);
        \coordinate (BBloo) at (  2, -8, -6);
        \coordinate (CCloo) at (  2,  8, -6);
        \coordinate (DDloo) at (  -6,  8, -6);



        \draw[dotted] (AAlo) -- (AAup);
        \draw[] (BBlo) -- (BBup);
        \draw[] (CClo) -- (CCup);
        \draw[] (DDlo) -- (DDup);

        \draw[] (AAupp) -- (BBupp);
        \draw[] (CCupp) -- (DDupp);

        \draw[dotted] (AAloo) -- (BBloo);
        \draw[] (CCloo) -- (DDloo);


        \coordinate (OAup) at (-19, -8,  6);
        \coordinate (OBup) at ( 12, -8,  6);
        \coordinate (OCup) at ( 12,  8,  6);
        \coordinate (ODup) at (-19,  8,  6);
        \coordinate (OAlo) at (-19, -8, -6);
        \coordinate (OBlo) at ( 12, -8, -6);
        \coordinate (OClo) at ( 12,  8, -6);
        \coordinate (ODlo) at (-19,  8, -6);

        \draw[] (OAup) -- (AAupp);
        \draw[] (DDupp) -- (ODup);
        \draw[] (ODup)  -- (OAup);

        \draw[dotted] (OAlo) -- (AAloo);
        \draw[] (DDloo) -- (ODlo);
        \draw[dotted] (ODlo)  -- (OAlo);

        \draw[] (BBupp) -- (OBup);
        \draw[] (OBup)  -- (OCup);
        \draw[] (OCup)  -- (CCupp);

        \draw[dotted] (OAup) -- (AAup);
        \draw[] (OCup) -- (CCup);
        \draw[] (ODup) -- (DDup);
        \draw[] (OBup) -- (BBup);

        \draw[] (OBlo) -- (OClo);
        \draw[] (BBlo) -- (OBlo);
        \draw[] (CClo) -- (OClo);

        \draw[dotted] (OBlo) -- (BBloo);
        \draw[] (OClo) -- (CCloo);
        \draw[dotted] (OAlo) -- (AAlo);
        \draw[] (ODlo) -- (DDlo);	

        \coordinate (R1) at ( -20.5, 8, 6);
        \coordinate (R2) at ( -20.5, 8, -6);
        \coordinate (R3) at ( -19, 8, 6);
        \coordinate (R4) at ( -19, 8, -6);
        \draw[<->] (R1)--(R2);
        \draw[] (R1)--(R3);
        \draw[] (R2)--(R4);
        \node at (-22.5,8,0) {$42D$};
        
        \coordinate (P1) at ( 2, 9.5, -6);
        \coordinate (P2) at (-19, 9.5, -6);
        \coordinate (P3) at ( 2, 8, -6);
        \coordinate (P4) at ( -19, 8, -6);
        \draw[<->] (P1)--(P2);
        \draw[] (P1)--(P3);
        \draw[] (P2)--(P4);
        \node at (-5.75,11.7,-6) {$42.5D$};
        
        \coordinate (Q1) at ( 12, 9.5, -6);
        \coordinate (Q2) at (  2, 9.5, -6);
        \coordinate (Q3) at ( 12, 8, -6);
        \coordinate (Q4) at ( 2, 8, -6);
        \draw[<->] (Q1)--(Q2);
        \draw[] (Q1)--(Q3);
        \draw[] (Q2)--(Q4);
        \node at (5.25,11.7,-6) {$20D$};
        
        \coordinate (S1) at ( 13.5,-8, -6);
        \coordinate (S2) at ( 13.5, 8, -6);
        \coordinate (S3) at ( 12,-8, -6);
        \coordinate (S4) at ( 12, 8, -6);
        \draw[<->] (S1)--(S2);
        \draw[] (S1)--(S3);
        \draw[] (S2)--(S4);
        \node at (14.7,0,-6) {$5D$};
        
        \coordinate (F1) at (27,0,0);
        \coordinate (F2) at (21,0,0);
        \draw [->,>=stealth] (F1)--(F2);
        \node at (26,0,1) {$U_{\infty}$};

        \coordinate (O0) at (21,0,-6);
        \coordinate (O1) at (18,0,-6);
        \coordinate (O2) at (21,0,-4);
        \coordinate (O3) at (21,2,-6);

        \node at (18,0,-6.5) {$x$};
        \node at (20.5,0,-4) {$y$};
        \node at (20.5,2,-6) {$z$};
        \draw [->] (O0)--(O1);
        \draw [->] (O0)--(O2);
        \draw [->] (O0)--(O3);
	\end{tikzpicture}	
\caption{Sketch of the computational domain for the BARC flow, with the reference system.}
\label{fig:sketch}
\end{figure}
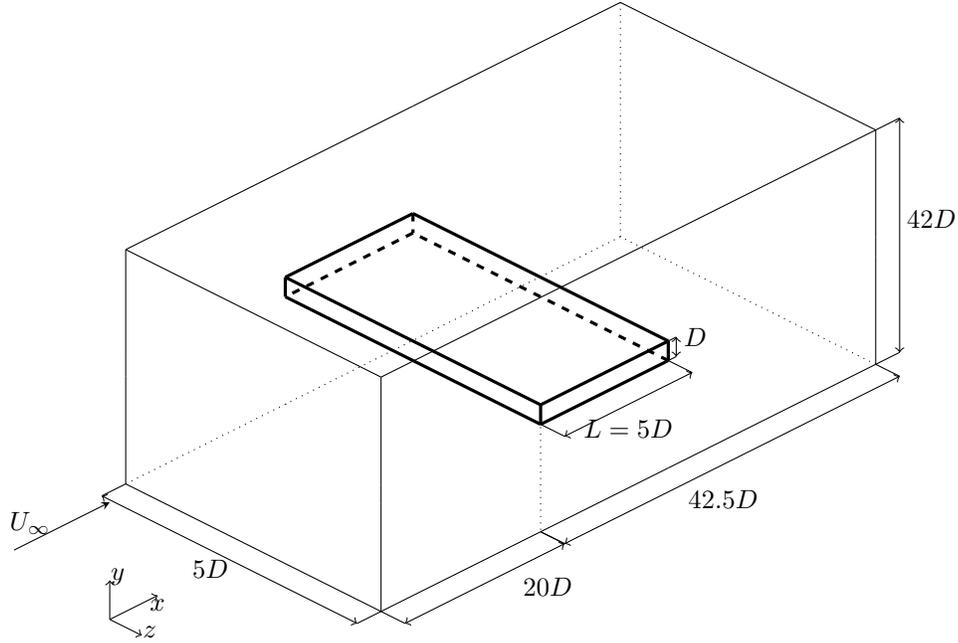

Figure \ref{fig:sketch} shows the geometry, the reference system and the notation. 
A Cartesian reference system is used with the origin placed at the LE of the cylinder. The cylinder has length $L$ and thickness $D$, curvature radius $R$ of its LE corners, and it is immersed in a uniform stream with velocity $U_\infty$ aligned with the $x$ direction. The $y$ and $z$ axes indicate the vertical and spanwise directions, respectively. The Reynolds number is based on the incoming velocity, the cylinder thickness and the kinematic viscosity $\nu$ and is set to $Re \equiv U_\infty D /\nu = 3000$ for all the considered cases.

The flow is governed by the incompressible Navier--Stokes equations:
\begin{equation}
\begin{aligned}
& \frac{\partial \vect{u}}{\partial t}  + \left( \vect{u} \cdot \vect{\nabla} \right) \vect{u} = - \vect{\nabla} p + \frac{1}{Re} \nabla^2 \vect{u} \\
& \vect{\nabla} \cdot \vect{u} = \vect{0}
\end{aligned}
\end{equation}
where $\vect{u}=(u,v,w)$ is the velocity vector and $p$ is the pressure. All quantities are made dimensionless with $U_{\infty}$ and $D$. The mean field is indicated with capital letters, i.e. $\vect{U}=(U,V,0)$ and $P$, while the fluctuations are indicated with a prime, $\vect{u'}=(u',v',w')$ and $p'$. The computational domain extends from $-20 \le x \le 42.5$, $-21 \le y \le 21$ and $-2.5 \le z \le 2.5$, with the cylinder placed at $0 \le x \le 5$, $-0.5 \le y \le 0.5$ and $-2.5 \le z \le 2.5$. No-slip and no-penetration conditions are imposed at the cylinder surface, the unperturbed velocity $\vect{u}=(U_{\infty},0,0)$ is assigned at the inlet and at the far field; periodic conditions are used at the spanwise boundaries to account for the spanwise homogeneity and a convective outlet condition $\partial \vect{u} / \partial t = U_{\infty} \partial \vect{u}/ \partial x$ is set at the outlet boundary.

The Navier--Stokes equations are solved using the DNS code introduced by \cite{luchini-2016} and already used for the BARC flow simulation in \cite{chiarini-quadrio-2021}. It solves the governing equations in primitive variables and employs second-order finite differences on a staggered grid. The cylinder is represented via an implicit, second-order accurate immersed-boundary method introduced in \cite{luchini-2013}. For further details on the numerical method see \cite{chiarini-quadrio-2021}, from which also the discretisation choices are derived. The number of points is $N_x=1776$, $N_y=942$ and $N_z=150$ in the three directions. An uniform distribution is employed in the spanwise direction, whereas a geometric progression is used for the streamwise and vertical directions to properly refine the flow region close to the LE and TE corners, where the spacing is $\Delta x = \Delta y \approx 0.0015$.

The present work describes the results of three new simulations carried out on purpose and compares them to the baseline results taken from \cite{chiarini-quadrio-2021}. Two cases are for the rounded LE corners and the third case deals with the analytical corner correction. The rounding is quantitatively defined by the radius $R$ of the inscribed quarter circle. Two curvature radii have been considered, namely $R/D=1/128$ and $R/D=1/64$, and denoted in the paper as cases C1 and C2. The number of points spanning the curvature radius are $6-7$ for case C1 and twice that for C2. The simulations are advanced in time using a varying time step to ensure that the Courant-Frederic-Levy CFL number remains at $CFL \le 1$, corresponding to an average value of the time step of $ \Delta t \approx 0.0013$. Before collecting statistics, every simulation is advanced in time long enough to reach a statistically-stationary state. For the simulations with rounded LE corners, statistics are collected over $300 D/U_\infty$ time units and $300$ flow fields are sampled at unitary time separation. For the simulation with the analytical corner correction, statistics are collected for a total time of $470 D/U_\infty$, again with a unitary sampling time. To increase the statistical sample, symmetries of the flow in the vertical direction are used. Thus, mean quantities, indicated by the operator $\aver{\cdot}$, are computed by averaging in time and exploiting both the homogeneity in the spanwise direction and the statistical symmetry in the vertical direction.

\subsection{The mean and instantaneous flow}
\label{sec:meanflow}

\begin{figure}
\centering
\includegraphics[trim=20 0 200 400,clip,width=1\textwidth]{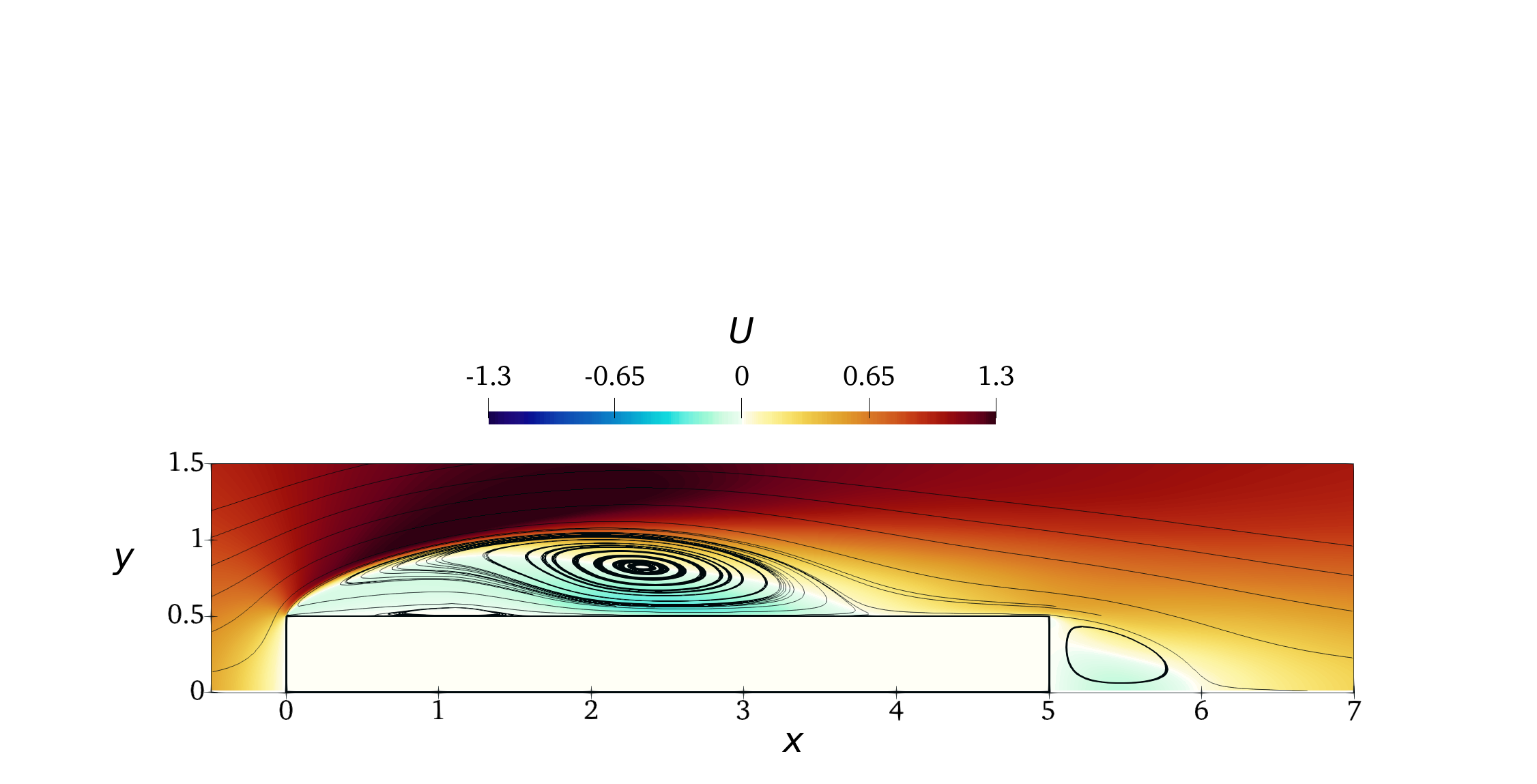}
\includegraphics[trim=20 0 200 400,clip,width=1\textwidth]{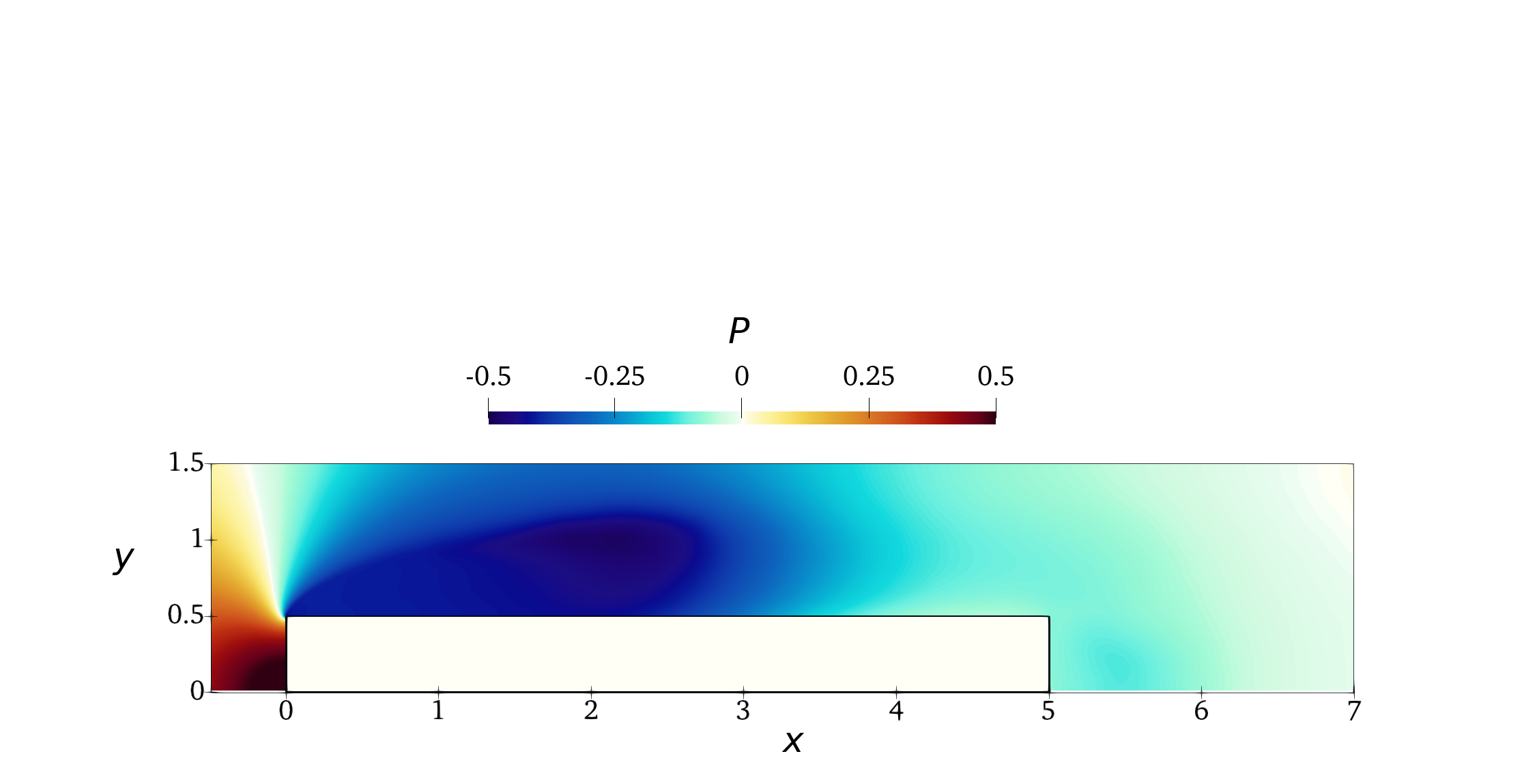}
\caption{Mean flow. Top: streamlines and color contour for the mean streamwise velocity component $U$. Bottom: mean pressure $P$.}
\label{fig:mean}
\end{figure}

To set the stage, the mean flow obtained with the reference sharp LE corners is briefly illustrated. For further details see \cite{chiarini-quadrio-2021}. Figure \ref{fig:mean} plots the mean streamlines superimposed on the map of $U$ in the top panel and the mean pressure $P$ in the bottom panel. A shear layer with negative vorticity starts from the sharp LE corner; the flow reattaches over the cylinder side before eventually separating again at the TE. Three areas of recirculation can be identified: two of them are above the longitudinal side of the cylinder and one is in the wake region. The first recirculating region is delimited by the shear layer separating at the LE and reattaching at $x_r \approx 3.955$, and it is hereafter referred to as the primary vortex. Its centre of rotation, i.e. the elliptical stagnation point with $U=V=0$, is placed at $(x,y) \approx (2.357,0.83)$. As shown in the bottom panel, the core of the primary vortex shows large negative values of pressure. Within the primary vortex, a smaller counter-rotating recirculating region is present, hereafter referred to as the secondary vortex. It is generated by the reverse boundary layer in the near-wall region of the primary vortex, which separates moving upstream owing to the adverse pressure gradient \cite{simpson-1989}. The secondary vortex extends for $0.63 \le x \le 1.59$ and its centre of rotation is placed at $(x,y) \approx (1.2,0.541)$. The third recirculating region in the wake, hereafter referred to as wake vortex, is delimited by the shear layer separating from the sharp TE. Its centre of rotation is placed at $(x,y) \approx (5.415,0.25)$ and extends up to $x \approx 5.947$.

\begin{figure}
\includegraphics[trim=150 0 560 0,clip,width=0.49\textwidth]{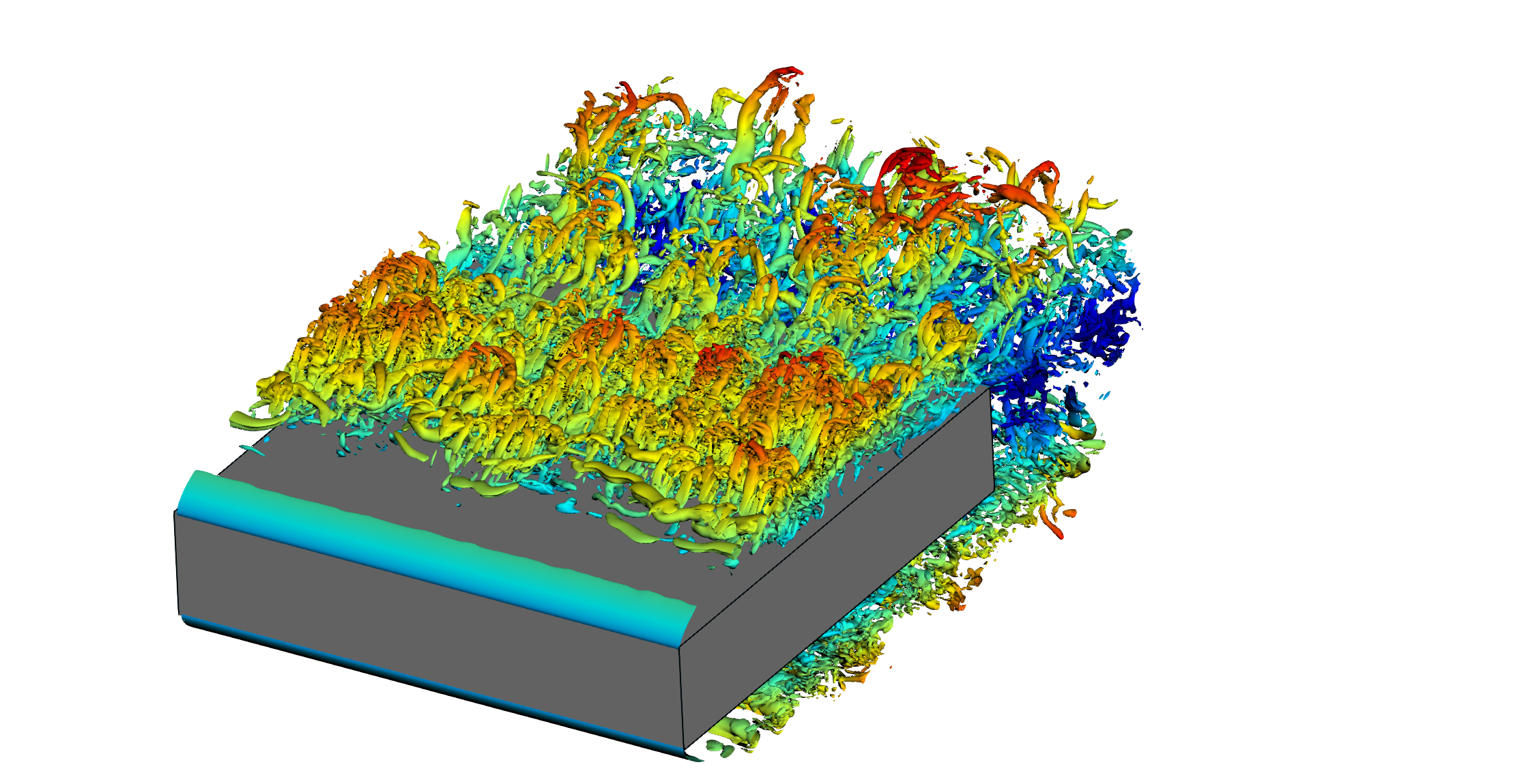}
\includegraphics[trim=150 0 560 0,clip,width=0.49\textwidth]{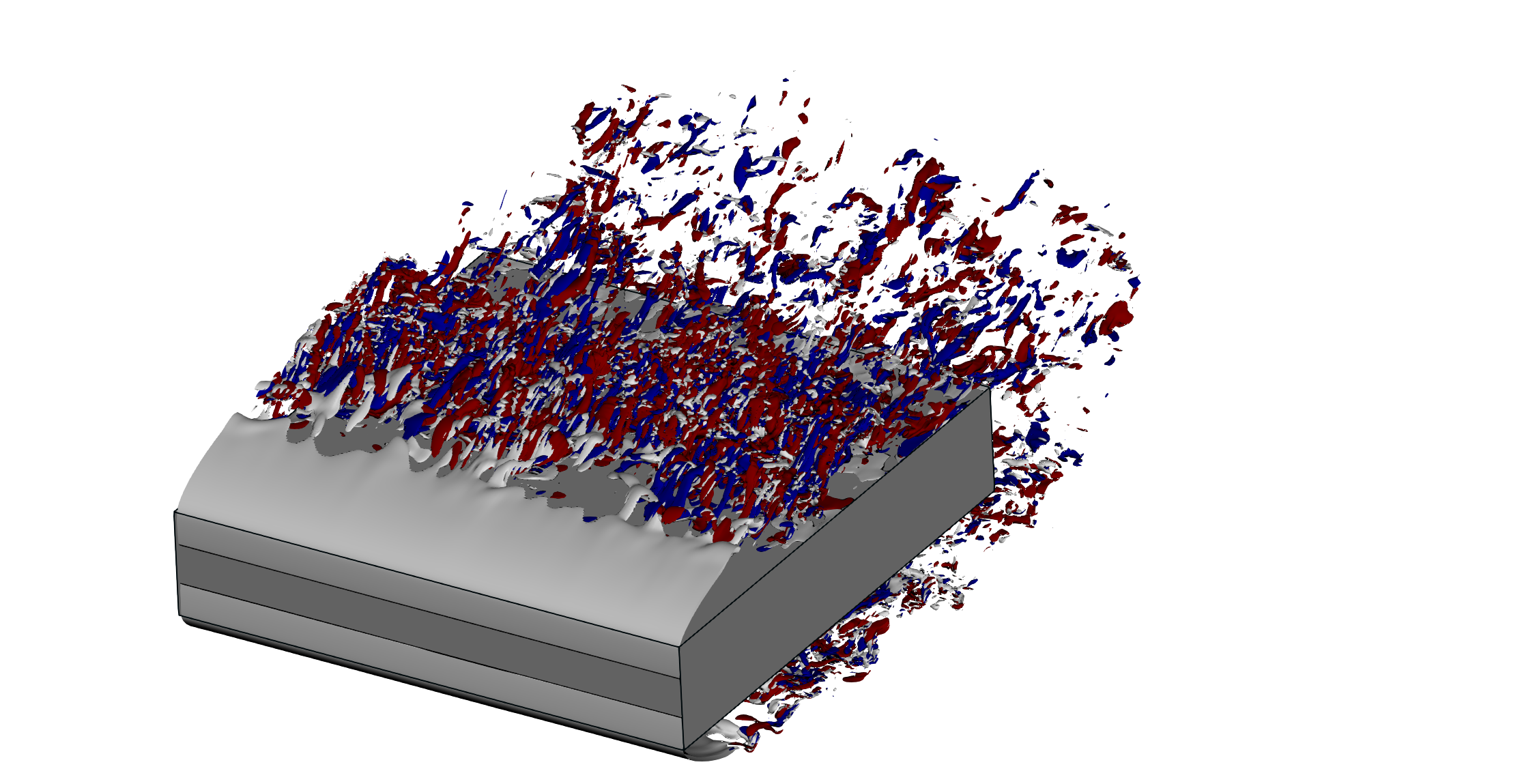}
\caption{Vortical structures in an instantaneous snapshot. Left: isosurface $\lambda_2=-10$ coloured with $|y|$; the blue-to-red colourmap goes from $|y|=0.5$ to $|y|=1.75$. Right: isosurfaces of $\omega_x=10$ (red), $\omega_x=-10$ (blue) and $|\omega_z|=17$ (grey).}
\label{fig:vortical_structures}
\end{figure}

Figure \ref{fig:vortical_structures} shows the turbulent structures populating an instantaneous snapshot of the BARC flow. They are visualised as isosurfaces of the second larger eigenvalue $\lambda_2$ of the velocity gradient tensor \cite{jeong-hussain-1995}. Contours of streamwise and spanwise vorticity $\omega_x$ and $\omega_z$ are also used to visualise the orientation of the structures.

After the LE separation, the flow remains initially laminar, until at $x \approx 0.5$ a Kelvin--Helmholtz instability of the shear layer occurs, which breaks it into large-scale spanwise tubes. Moving downstream the tubes are stretched by the mean gradient and roll up, originating hairpin-like structures. Further downstream a complete transition to turbulence is observed. The hairpin vortices break down into elongated streamwise vortices that are easily visualised by the positive and negative contours of $\omega_x$. In this region of the flow the large-scale motions derived from the Kelvin--Helmholtz instability coexist with the small-scales structures associated with the turbulent motions. Finally, at the TE the flow separates again and the turbulent structures are convected in the turbulent wake.

\section{Rounding the LE corners}

This section discusses the effects of rounding the LE corners. Two rather small curvature radii characterize cases C1 and C2, since the aim of the present work is to investigate the effect of manufacturing imperfections on the BARC flow.  

\subsection{The mean flow}

The mean flow with rounded corners closely resembles the one of the reference configuration with sharp corners \cite{chiarini-quadrio-2021}, with only small changes in the size of the three recirculating regions and in the position of their centre of rotation. These changes are summarised in table \ref{tab:diff-meanflow}, where the extent of the three recirculating regions and the position of their centre of rotation are reported.
\begin{table}
\caption{Comparison of size and positions of the three recirculating regions, for the cases with rounded corners and the reference one with sharp corners from Ref. \cite{chiarini-quadrio-2021}. $x_s$ and $x_e$ are the start and end coordinates, and $L$ indicates the length of each region, whereas $x_c$ and $y_c$ are the coordinates of their centre of rotation.}
\label{tab:diff-meanflow}
\centering
\begin{tabular}{ccccc}
 \hline
                                   &             & sharp          & $C1$     & $C2$         \\
 \cline{1-5}                                  
 \multirow{4}{*}{Primary vortex}   & $x_{s,1}$        & $0$           & $ 0.005$      & $0.01$           \\
                                   & $x_{e,1}$        & $3.955$       & $ 3.895$      & $3.89$           \\
                                   & $L_1$        & $3.955$       & $ 3.89$       & $3.88$           \\
                                   & $(x_c,y_c)$ & $(2.357,0.83)$& $(2.361,0.81)$& $(2.43,0.81)$    \\
 \cline{1-5}                                  
 \multirow{4}{*}{Secondary vortex} & $x_{s,2}$       & $0.63$        &  $0.75$        & $0.97$           \\
                                   & $x_{e,2}$       & $1.59$        &  $1.585$       & $1.75$           \\
                                   & $L_2$       & $0.96$        &  $0.835$       & $0.78$           \\
                                   & $(x_c,y_c)$ & $(1.2,0.541)$ &  $(1.17,0.531)$ & $(1.315,0.533)$ \\
 \cline{1-5}                                  
 \multirow{4}{*}{Wake vortex}      & $x_{s,3}$       & $5$            & $5$           & $5$              \\ 
                                   & $x_{e,3}$       & $5.975$        & $6.01$        & $6.02$           \\
                                   & $L_3$       & $0.975$        & $1.01$        & $1.02$           \\
                                   & $(x_c,y_c)$ & $(5.415,0.25)$ & $(5.425,0.25$ & $(5.43,0.25$)    \\
  \hline
\end{tabular}
\end{table}

\begin{figure}
\centering
\includegraphics[width=0.8\textwidth]{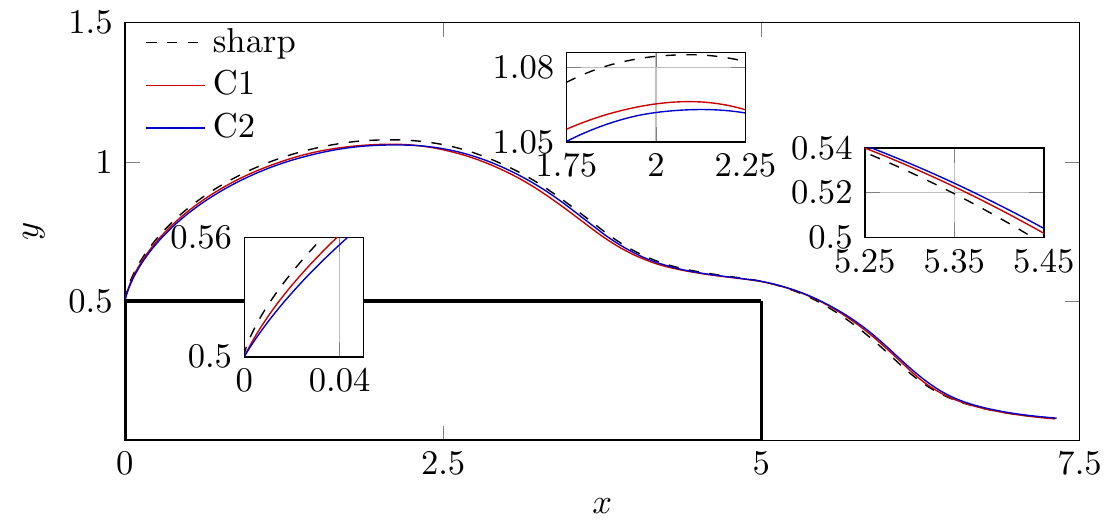}
\caption{Mean streamline passing through the point $(x,y)=(0,0.5001)$, for the sharp and rounded configurations. The zoomed insets highlight three regions close to the LE, close to the top of the primary recirculation, and in the near wake.}
\label{fig:Line}
\end{figure}

How rounding affects the primary and wake vortices is shown by the streamline passing near the sharp LE, shown in figure \ref{fig:Line}. The starting point of the streamlines is placed just above the LE corner at $(x,y)=(0,0.5001)$ for both the sharp and rounded configurations, although in the latter cases the actual separating point is slightly shifted downstream; see table \ref{tab:diff-meanflow}. The differences shown in figure \ref{fig:Line} have been verified to be robust to small shifts of the seeding point.
This streamline delimits first the primary vortex and then, after passing over the trailing edge, the wake vortex. Close to the separation point the line has a lower slope in the rounded cases, indicating a lower inclination of the shear layer. At $x \approx 2$ the streamline shows that the vertical extent of the primary vortex decreases for increasing values of $R$. Finally, when the streamline crosses the TE a milder slope develops, consistently with the smaller extension of the wake vortex reported in table \ref{tab:diff-meanflow}. 

\begin{figure}
\centering
\includegraphics[width=0.9\textwidth]{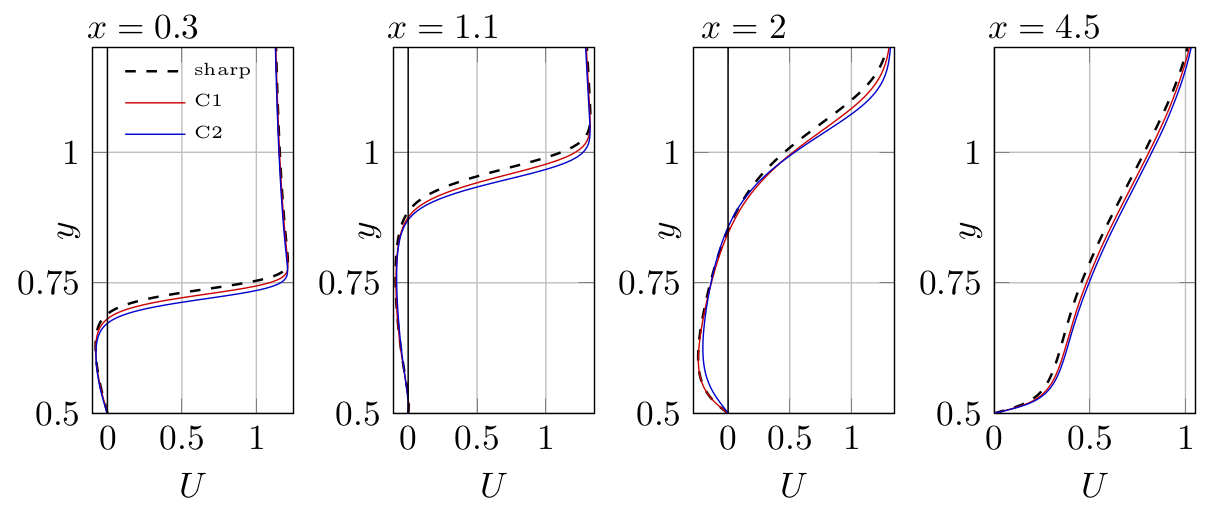}
\caption{Vertical profiles of the streamwise component of the mean velocity $U$ at four different stations over the cylinder side.}
\label{fig:U_vel_profiles}
\end{figure}

Figure \ref{fig:U_vel_profiles} presents the vertical profile of the $U$ velocity component at different streamwise locations over the cylinder side, i.e. $x = 0.3, 1.1, 2, 4.5$.
The first three panels describe the changes within the primary vortex. As shown in figure \ref{fig:Line}, the rounded corners lead to a slightly reduced vertical extent of the primary vortex, as the shear layer separates from the LE with a lower angle. This is also conveniently visualised by the coordinate $y_{U=0}$ where the mean streamwise velocity component becomes zero. In the first portion of the primary vortex $y_{U=0}$ is shifted towards the wall in the rounded configurations; see the two left panels of figure \ref{fig:U_vel_profiles}. Moving downstream for $x \ge 1.5$, instead, the difference is less evident indicating that the main changes are localised close to the LE; see the third panel where $y_{U=0}$ is almost the same for the sharp and rounded configurations. A decrease of the shear-layer separation angle is consistent with a lower longitudinal size of the primary vortex, which is seen in table \ref{tab:diff-meanflow} to decrease from $L_1 = 3.95$ to $L_1 = 3.89$ for case C1 and $L_1 = 3.88$ for C2, i.e. approximately $2\%$. A further effect of the rounding is the decrease of the backflow in the core of the primary vortex shown at $x=2$, where $U$ is less negative in the rounded configurations. This is consistent with the results of \cite{lamballais-etal-2010} who analysed the effect of (large) LE roundings on infinite D-shaped bodies. Interestingly, the rounded corners also affect the mean field after the reattachment, as seen in the last panel of figure \ref{fig:U_vel_profiles} at $x=4.5$, where the mean flow is accelerated. This is explained by the slightly lower extension of the primary vortex that enables a larger development of the successive boundary layer before its separation, and is associated to an enhancement of the turbulent activity as shown in the following section. 

Table \ref{tab:diff-meanflow} quantifies the modifications of the three main vortices. When the LE corners are rounded, the upstream separation point slightly moves downstream  and is found close to the end of the curvature, where a second-derivative discontinuity takes place. we found the primary vortex to start at $x_{s,1}=0.005$ for C1 and at $x_{s,1}=0.01$ for C2. As already mentioned, this small downstream shift together with the decrease of the shear layer separating angle leads to a decrease of the longitudinal and vertical extensions of the primary vortex. The same effect has been found in preliminary two- and three-dimensional simulations (not shown here) in the laminar regime at $Re=500$, with $ 1/128 \le R/D \le 1/2$. The centre of rotation of the primary vortex slightly moves towards the reattachment point as $R$ increases, consistently with the results of \cite{lamballais-etal-2010}. Rounding the corners also affects the smaller secondary vortex. Indeed, for both C1 and C2 we have found that its longitudinal size consistently decreases up to $L_2 \approx 0.835$ ($\approx -12 \%$) for C1 and $L_2 \approx 0.78$ ($\approx -18\%$) for C2. However, in the two rounded cases there are some differences. For C1 the position of the centre of rotation is almost unchanged, since $x_{s,2}$ moves downstream and $x_{e,2}$ moves upstream. For C2, instead, the decreased size of the recirculating vortex is accompanied by an overall downstream shift, as indicated by the positions of $x_{s,2}$, $x_{e,2}$ and of the centre of rotation. Interestingly, the LE rounding also affects the size of the wake vortex. Indeed, by increasing the curvature radius its length slightly increases up to $L_3 \approx 1.01$ for C1 ($\approx +3.5 \%$) and up to $L_3 \approx 1.02$ for C2 ($\approx +4.5 \%$). Such modifications are consistent with the acceleration of the flow in the last part of the cylinder side, resulting in a shear layer that separates from the sharp TE with larger velocity.

\subsection{Turbulent kinetic energy}
\begin{figure}
\centering
\includegraphics[trim=20 0 200 400,clip,width=1\textwidth]{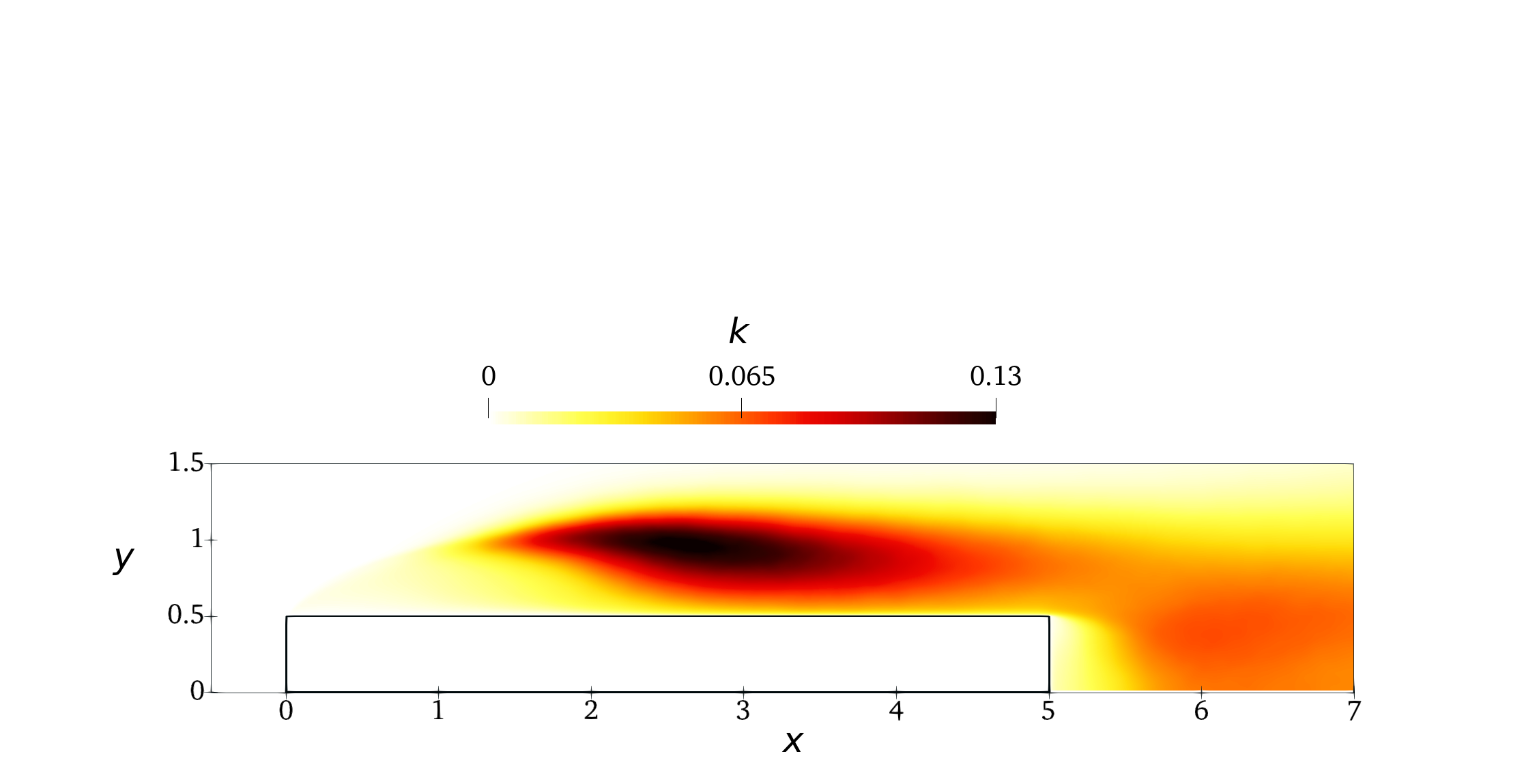}
\caption{Map of the turbulent kinetic energy for the sharp configuration.}
\label{fig:map_k}
\end{figure}
\begin{figure}
\centering
\includegraphics[width=0.9\textwidth]{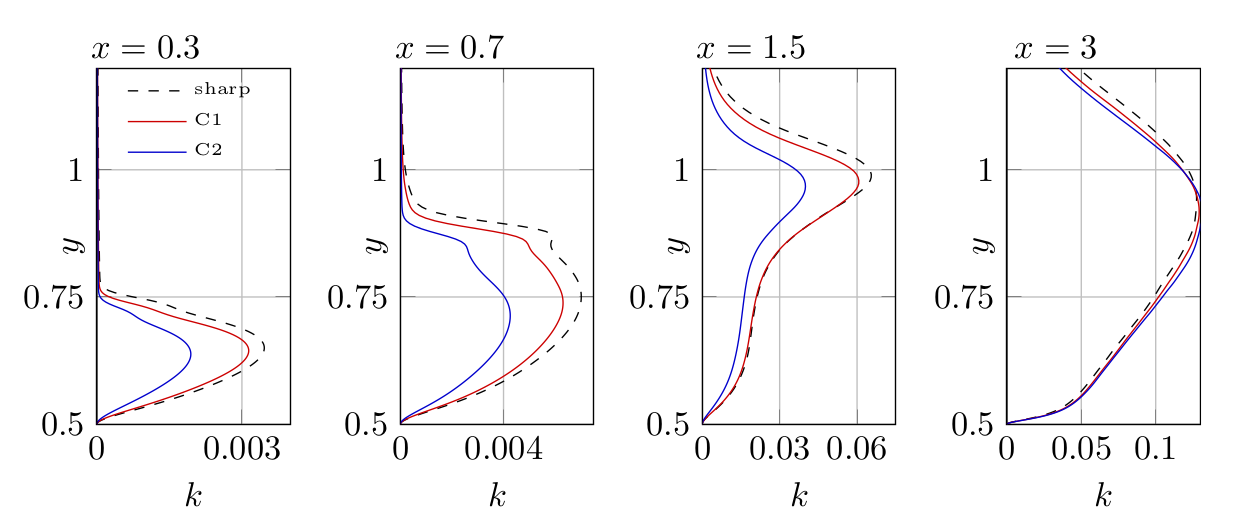}
\caption{Vertical profiles of the turbulent kinetic energy $k$ at four different stations over the cylinder side, i.e. $x=0.3,0.7,1.5,3$.}
\label{fig:k_vel_profiles}
\end{figure}

Rounding the corners also affects the fluctuating velocity field. Figure \ref{fig:map_k} plots the turbulent kinetic energy $k=\aver{u_i' u_i'}/2$ (repeated index implies summation) for the sharp configuration; figure \ref{fig:k_vel_profiles} plots vertical profiles of $k$ at different streamwise locations, i.e. $x=0.3, 0.7, 1.5, 3$, for the sharp and rounded cases. Close to the LE, i.e. for $x < 1$, $k$ is very small, confirming the almost laminar flow state. Moving downstream, $k$ quickly increases indicating a sharp transition to the turbulent state, as can be seen by inspecting an instantaneous velocity field in figure \ref{fig:vortical_structures}. The turbulent activity is most intense in the core of the primary vortex as indicated by the maximum of $k$ found at $(x,y) \approx (2.7,0.96)$; however large values of $k$ are observed also in the near-wake region close to the TE as shown by the presence of an additional local peak at $(x,y) \approx (6.15,0.36)$.

For C1 the vertical profiles of $k$ are very close to those of the sharp configuration. This further confirms that -- at least at this Reynolds number -- a small rounding of the LE corners does not lead to an abrupt change of the flow topology. In the rounded configuration the intensity of the velocity fluctuations decreases only near the LE: it seems that the rounded corners lead to a spatial delay of the development of the velocity fluctuations, in agreement with the results of \cite{rocchio-etal-2020,lamballais-etal-2008,lamballais-etal-2010,cimarelli-franciolini-crivellini-2020}. Close to the LE, the profiles of $k$ show a well-defined peak in correspondence of the shear layer (see the first three panels of figure \ref{fig:k_vel_profiles}). This agrees with the notion that, close to the LE, the fluctuations are mainly generated by the Kelvin-Helmholtz instability of the shear layer. Moving downstream, $k$ becomes distributed over a wider range of $y$, until a completely turbulent state is reached and large values of $k$ are observed in the overall extension of the primary vortex, revealing the presence of other production mechanisms. Interestingly, for $x \ge 2.5$ the intensity of $k$ for $y \le 1$ is larger in the rounded configurations, consistently with the picture of a delayed development of the turbulent fluctuations; see the right panel of figure \ref{fig:k_vel_profiles}. This accompanies the increased $U$ observed in this region of the flow. Moreover, in the rounded cases, owing to the reduced vertical extent of the primary vortex, $k$ drops to almost zero at lower $y$ compared to the sharp configuration.

The delay in the development of the turbulent kinetic energy in the rounded configurations and the larger $k$ observed after the reattachment point, may be explained by the differences in the production $P_k$ and dissipation $\epsilon_k$ in its budget. These terms read
\begin{equation}
P_k = - \aver{u'u'} \frac{\partial U}{\partial x} -  \aver{v'v'} \frac{\partial V}{\partial y} -  \aver{u'v'} \left( \frac{\partial U}{\partial y} + \frac{\partial V}{\partial x} \right),
\end{equation}
\begin{equation}
\epsilon_k = \nu \aver{\frac{\partial u'}{\partial x_j} \frac{\partial u'}{\partial x_j} } +
             \nu \aver{\frac{\partial v'}{\partial x_j} \frac{\partial v'}{\partial x_j} } +
             \nu \aver{\frac{\partial w'}{\partial x_j} \frac{\partial w'}{\partial x_j} },
\end{equation}
where repeated indices imply summation.

\begin{figure}
\centering
\includegraphics[trim=20 0 200 400,clip,width=1\textwidth]{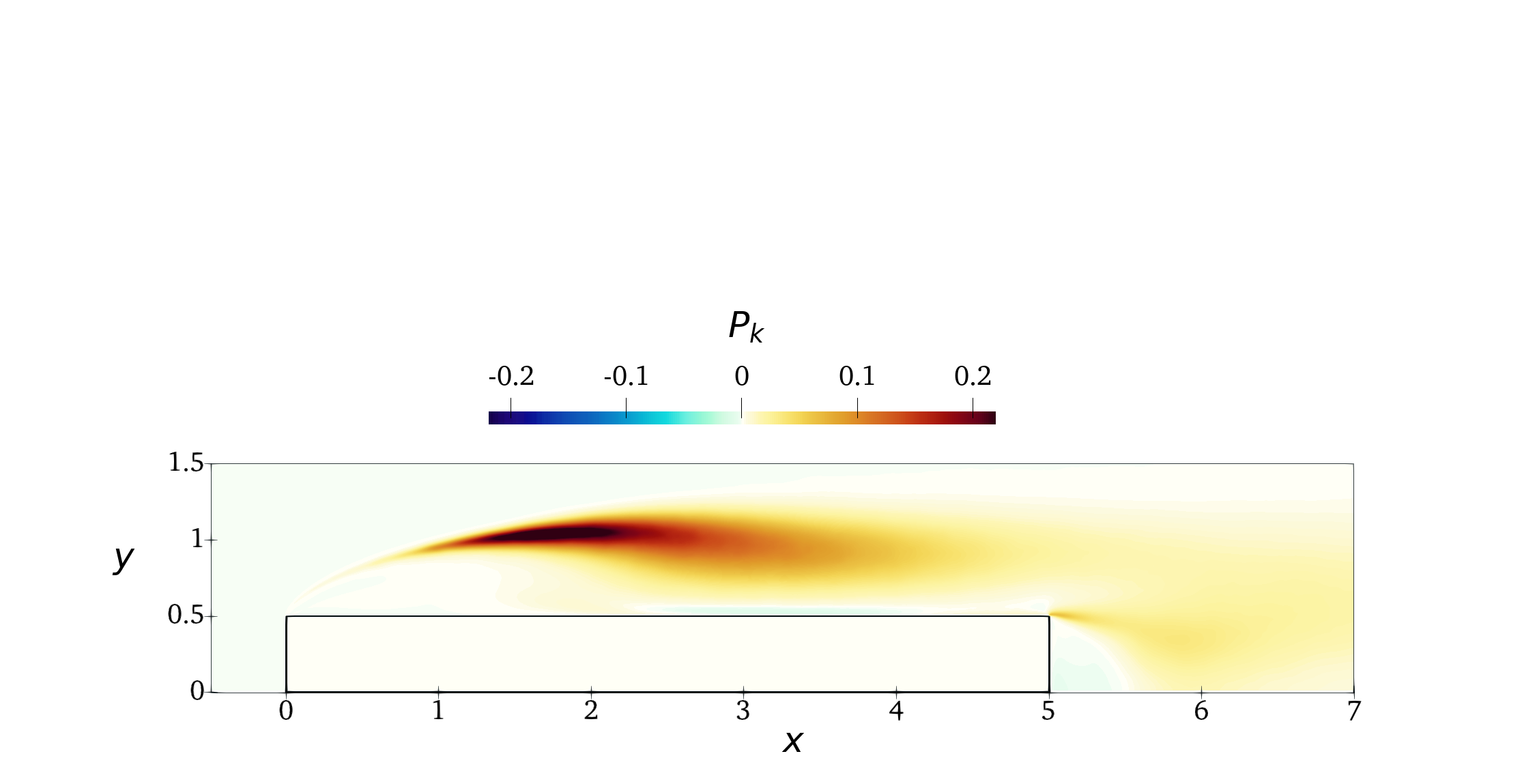}
\caption{Map of the production term $P_k$ of the turbulent kinetic energy for the sharp configuration.}
\label{fig:map_P}
\end{figure}

\begin{figure}
\centering
\includegraphics[width=0.9\textwidth]{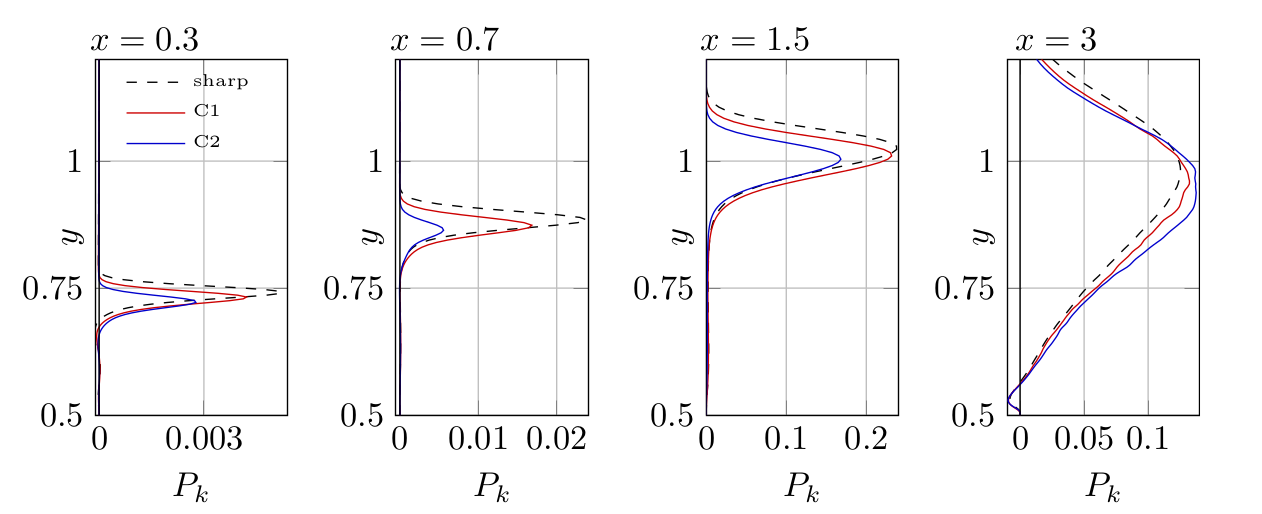}
\caption{Vertical profiles of $P_k$ at four different station over the cylinder side as in figure \ref{fig:k_vel_profiles}.}
\label{fig:Prodk_vel_profiles}
\end{figure}

Figure \ref{fig:map_P} plots the map of the production for the reference sharp configuration; figure \ref{fig:Prodk_vel_profiles}, instead, plots vertical profiles of $P_k$ at different streamwise positions over the cylinder side, i.e. $x=0.3,0.7,1.5,3$, for the sharp and rounded configurations.
For $x \le 0.5$ low values of $P_k$ are observed, which is consistent with the low values of $k$ in figure \ref{fig:map_k}. Along the shear layer $P_k$ is positive, but it becomes negative just below. However, as already discussed in \cite{chiarini-quadrio-2021}, this negative $P_k$ does not correspond to the negative production rate of $k$ localised in the shear layer seen by \cite{cimarelli-etal-2019-negative}. Moving downstream at $x \approx 1.3$, $P_k$ peaks in correspondence of the shear layer. This implies that, when the Kelvin--Helmholtz instability takes place, energy is being drained from the mean flow to feed the fluctuating field. Further downstream, large values are observed also at lower $y$ in the core of the primary vortex indicating that a further production mechanism different from the Kelvin--Helmholtz instability of the shear layer is occurring. Arguably, this is associated to the interaction between the streamwise-aligned vortices observed in figure \ref{fig:vortical_structures} to populate this region of the flow and the large mean velocity gradients. For $x \ge 2.5$ a region with slightly negative $P_k$ is observed in the vicinity of the cylinder side, indicating that energy feeds back from the fluctuations to the mean field. Thus, despite the presence of the wall, the mechanism sustaining the turbulent fluctuations close to the longitudinal side of the cylinder differs from what observed in the canonical wall-bounded flows, where $P_k$ is always positive. Finally large positive values of $P_k$ are found in correspondence of the shear layer separating from the TE, again due to the larger mean velocity gradients.

Figure \ref{fig:Prodk_vel_profiles} show that, close to the LE, the production decreases when the corners are rounded. More downstream, as already observed for $k$, the differences between the three cases shrink, indicating that the production of fluctuating energy in the rounded cases becomes close to what observed in the sharp configuration. Rounding the LE corners thus leads to a downstream shift of the Kelvin--Helmholtz shear layer instability and to a delayed streamwise development of the velocity fluctuations. As also observed in \cite{lamballais-etal-2010}, this may explain the downstream shift of the centre of rotation of the primary vortex. Indeed, the shift of the instability source turns into an extended low-velocity region in the upstream part of the primary vortex that is arguably responsible of the downstream shift of the stagnation point. We note again how for increasing curvature radius the wall distance at which $P_k$ drops to zero becomes smaller, owing to the decreasing vertical extent of the primary vortex. Interestingly, for $x \ge 2.5$ the production term of the rounded cases overcomes that of the sharp configuration for $y \le 1$, indicating that in the second part of the longitudinal cylinder side the production of the fluctuating energy is enhanced by the rounded corners. This agrees with the slight increase of $k$ in the same region and with the picture of a stronger shear layer separating from the TE.

\begin{figure}
\centering
\includegraphics[trim=20 0 200 400,clip,width=1\textwidth]{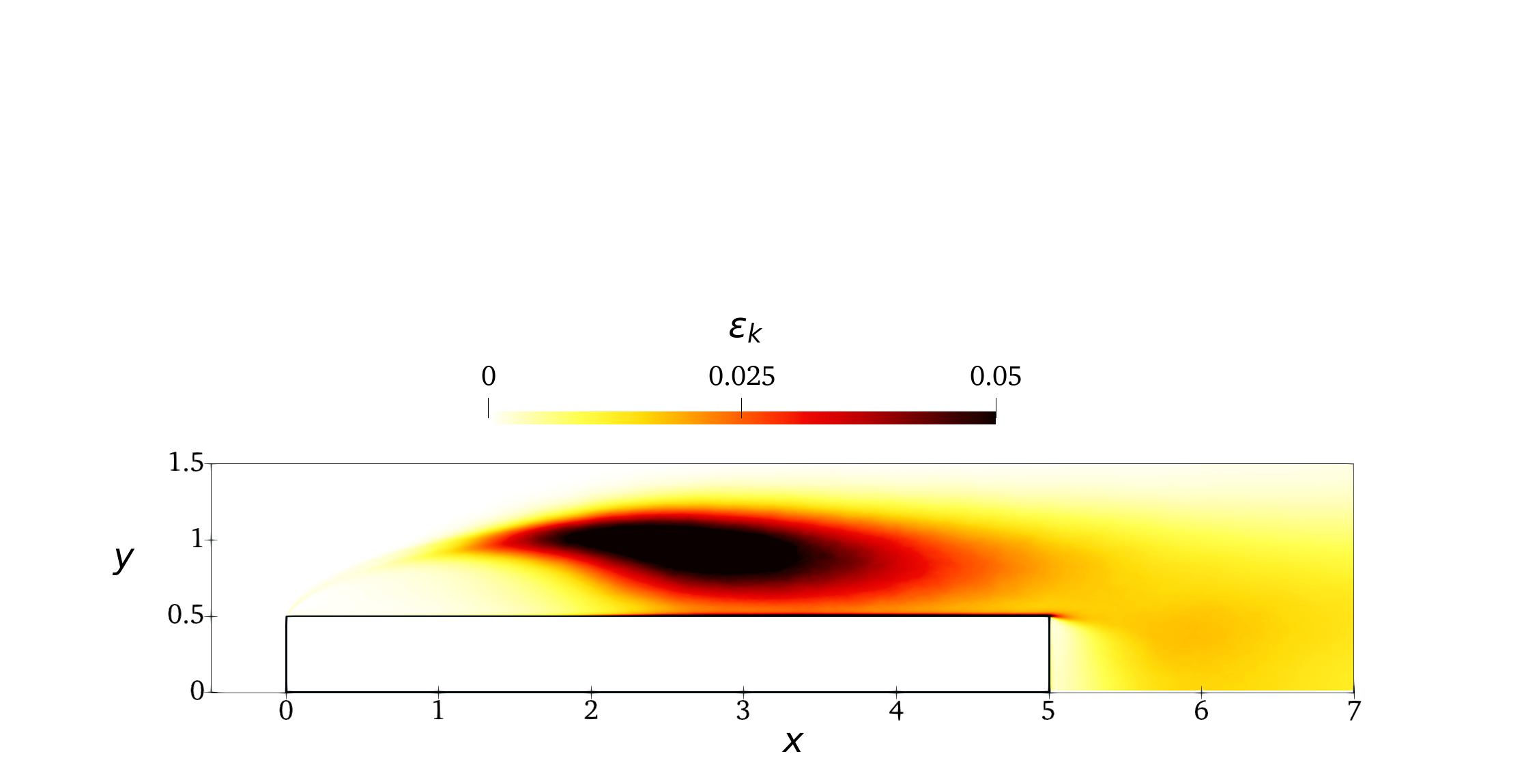}
\caption{Map of the dissipation rate of the turbulent kinetic energy}
\label{fig:map_epsilon}
\end{figure}

\begin{figure}
\centering
\includegraphics[width=0.9\textwidth]{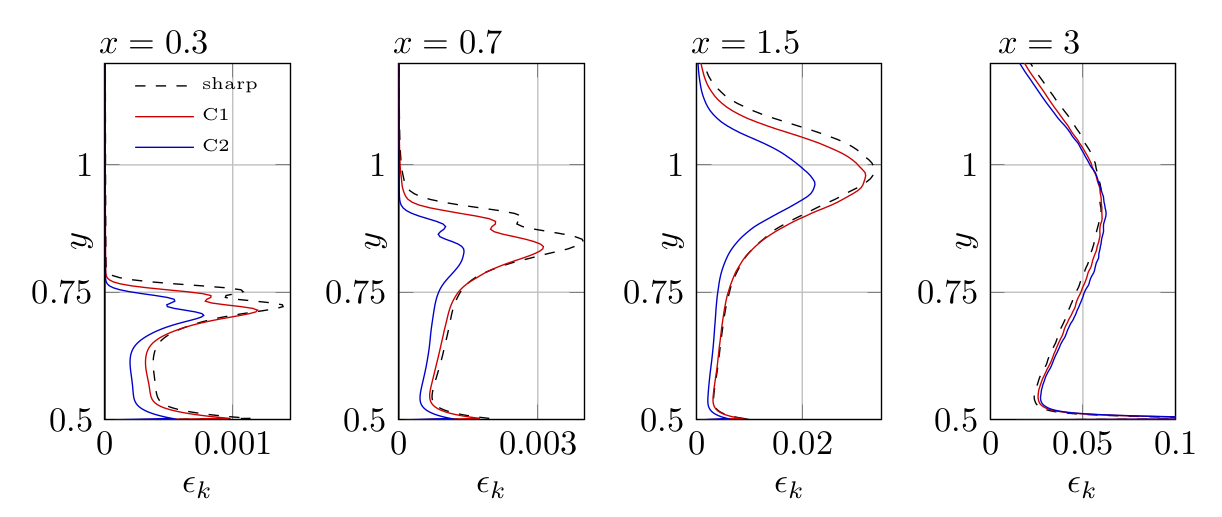}
\caption{Vertical profiles of the dissipation rate of the turbulent kinetic energy at four different stations over the cylinder side.}
\label{fig:Dissk_vel_profiles}
\end{figure}

\begin{figure}
\centering
\includegraphics[trim=20 0 200 400,clip,width=\textwidth]{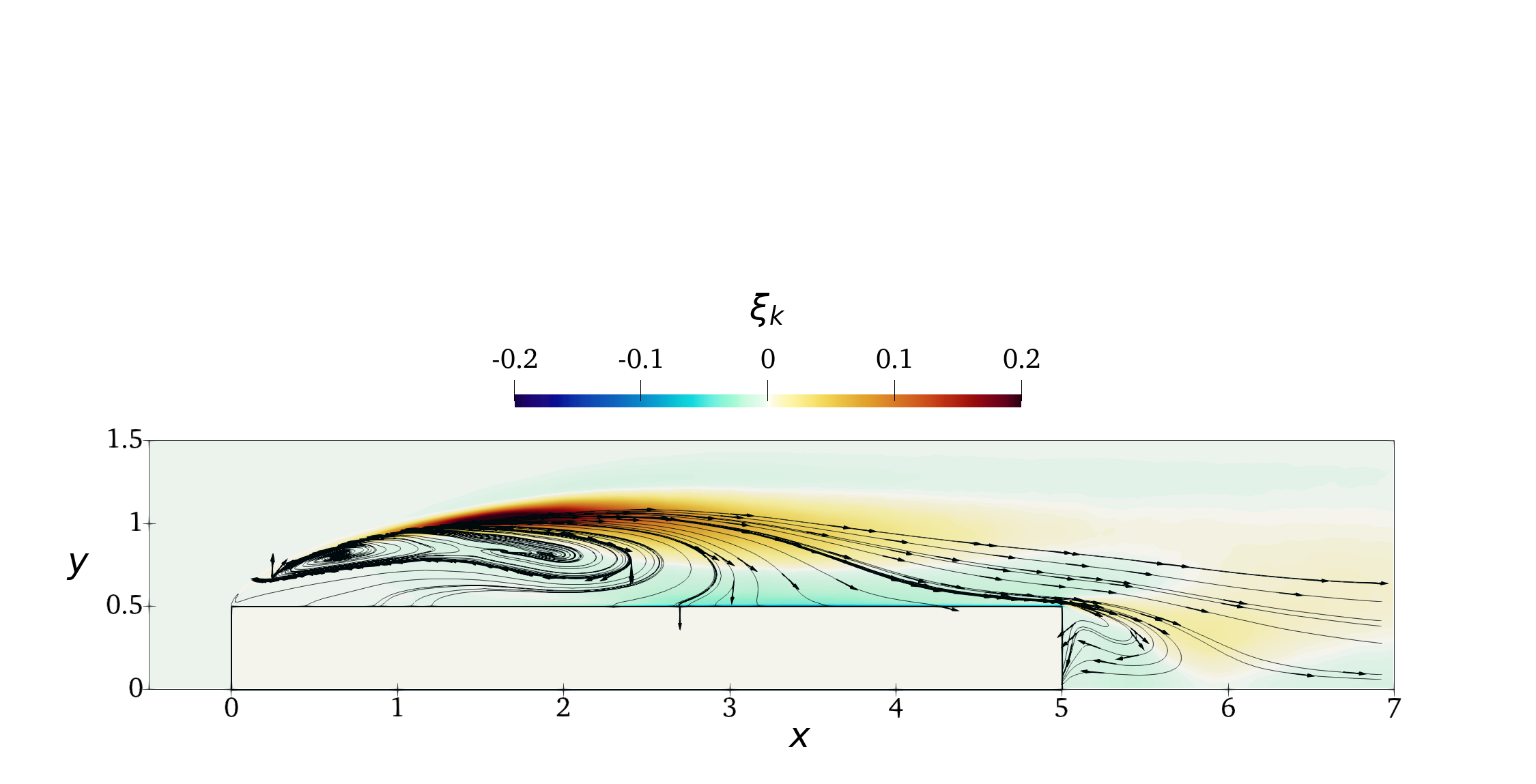}
\caption{Field lines of the flux vector $\boldsymbol{\psi}$ and colormap of the source $\xi_k=P_k - \epsilon_k$ for the reference sharp configuration. The arrows are tangent to the flux vector $\boldsymbol{\psi}$ and provide directional information.}
\label{fig:fluxes}
\end{figure}

Figures \ref{fig:map_epsilon} and \ref{fig:Dissk_vel_profiles} describe the dissipation $\epsilon_k$. Figure \ref{fig:map_epsilon} plots the map of $\epsilon_k$ for the sharp configuration, while figure \ref{fig:Dissk_vel_profiles} plots vertical profiles of $\epsilon_k$ for the three considered cases at four different streamwise locations, i.e. $x=0.3,0.7,1.5,3$.
Large values of $\epsilon_k$ occur in the shear layer for $x \ge 1$, in the core of the primary vortex and close to the cylinder side for $x \ge 2.5$, where viscous effects dominante. For $x < 1$, instead, the values of $\epsilon_k$ are much lower as the turbulence activity is locally scarce compared to the rest of the domain. As for production, the main differences between the rounded and sharp configurations are observed for $x <2.5$ where rounding leads to a decrease of $\epsilon_k$ in the overall extent of the primary vortex, in agreement with a picture of lower turbulent activity. Moving downstream, instead, the differences among the three profiles strongly decrease until for $x \ge 2.5$ $\epsilon_k$ becomes slightly larger in the rounded configurations for $y \le 1$. Note however that here the increase of $P_k$ in the rounded cases is larger than the increase of $\epsilon_k$. Therefore, this results in a net increase of the source term $\xi_k=P_k - \epsilon_k$ that determines an intensification of the spatial transports of $k$. Although there is no direct link, this may explain at least partially the larger values of $k$ observed in figure \ref{fig:k_vel_profiles} for the rounded cases. The spatial transport of $k$ is visualised by the two-dimensional flux vector $\boldsymbol{\psi}$ defined as \cite{pope-2000}:
\begin{equation}
\psi_j = \underbrace{ \frac{1}{2} \aver{u_i' u_i' u_j'} }_{\text{turbulent transport}} +
         \underbrace{ \aver{ p' u_j' } }_{\text{pressure transport}} +
         \underbrace{ U_j \frac{1}{2} \aver{u_i' u_i'} }_{\text{mean transport}} - 
         \underbrace{ \frac{\nu}{2} \frac{\partial}{\partial x_j} \aver{u_i' u_i'} }_{\text{viscous diffusion}} \ \ \text{with} \ j=x,y
\label{eq:flux_vector}
\end{equation}  
where repeated indices imply summation.
Note that this is half of the sum of the flux vector for the three normal stresses discussed in \cite{chiarini-quadrio-2021}. Figure \ref{fig:fluxes} plots the field lines of the flux vector over the colour contour of the source term $\xi_k$, or equivalently $\boldsymbol{\nabla} \cdot \boldsymbol{\psi}$. The fluxes allow a precise description of the spatial transfer, and their field lines visualise how the kinetic energy is transferred in space. Therefore, the fluxes explain the different positions at which $k$ and $\xi_k$ have their peak. Their divergence, $\boldsymbol{\nabla} \cdot \boldsymbol{\psi}$, provides quantitative information about the energetic relevance of the fluxes. When $\boldsymbol{\nabla} \cdot \boldsymbol{\psi}$ is positive, i.e. $\xi_k>0$, the fluxes are energised by local production mechanism. In contrast, when $\boldsymbol{\nabla} \cdot \boldsymbol{\psi}$ is negative the fluxes release energy to sustain locally the fluctuations. The flux lines originate where $\boldsymbol{\nabla} \cdot \boldsymbol{\psi}$ has large positive values and vanish where $\boldsymbol{\nabla} \cdot \boldsymbol{\psi}$ is negative. 

The large contribution of the Kelvin--Helmholtz instability to $P_k$ yields large $\xi_k>0$ along the shear layer with a peak at $(x,y) \approx (1.7,1)$; as expected the largest values are slightly shifted downstream in the rounded cases. Close to the cylinder side, i.e. approximately for $y \le 0.75$, the negative contribution of $\epsilon_k$ dominates and leads to $\xi_k<0$. A further region of positive $\xi_k$ is observed in correspondence of the TE shear layer. The fluxes of $k$ differ from the fluxes of the three normal stresses. The large values of $\xi_k$ in correspondence of the shear layer energise all the fluxes that in turn transfer the excess of $k$ over the entire domain. Four different types of lines are observed, which relate to different transport mechanisms. Some lines pass over the TE at relatively large $y$ and, dominated by the mean transport, continue in the wake region. A second group of lines passes very close to the TE and then they are further energised by the positive $\xi_k$ in the separating shear layer. These lines then vanish in correspondence of the rear vertical side of the cylinder, after having released $k$ within the wake vortex. These two groups of field lines indicate that the flow over the cylinder side influences both the wake vortex and the downstream wake. 

The other two types of lines remain confined within the primary vortex, pointing to a self-sustaining mechanism, as predicted by \cite{cimarelli-leonforte-angeli-2018}. Some are attracted by the cylinder side in the whole range $ 0 \le x \le 5$, indicating that part of the excess of $k$ produced by the Kelvin--Helmholtz instability is released near the wall where it is partially dissipated by viscous effects and partially feeds the mean flow (recall the negative values of $P_k$). The last type of lines shows a spiral-like behaviour in the upstream part of the cylinder side. Such lines indicate that part of the $k$ produced by the Kelvin--Helmholtz instability is transported and released upstream to feed the flow region close to the LE. This spiral-like pattern has two singularity points the lines are attracted to. For the sharp configuration they are located at $(x,y) \approx (0.6,0.81)$ and $(x,y) \approx (1.8,0.85)$. For the rounded configuration C1 the qualitative differences on these transfers are small, confirming again that a small rounding does not lead to an abrupt change in the transport of $k$. However, for C2 differences are more significant. In particular, the spiralling pattern is shifted downstream: the upstream singularity point moves to $(x,y) \approx (0.87,0.83)$. Overall, this means that in the rounded configurations the downstream shift of the Kelvin--Helmholtz instability is accompanied by a downstream shift of the transfer mechanism involving the upstream part of the primary vortex. Therefore, both effects are responsible of the decrease of the intensity of $k$ in the region close to the LE.

\begin{figure}
\centering
\includegraphics[width=0.9\textwidth]{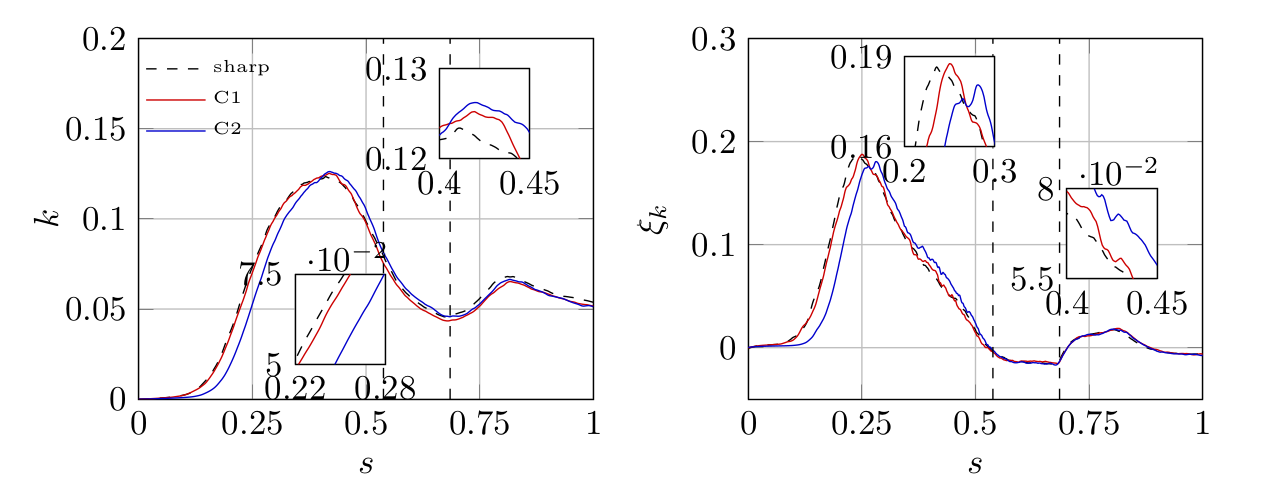}
\caption{Evolution of $k$ (left) and $\xi_k$ (right) over the mean streamline originating at $(x,y)=(0,0.5)$ shown in figure \ref{fig:Line} as a function of the normalised curvilinear coordinate $s$. The left vertical dashed line marks the reattachment point, and the right dashed line marks the TE.}
\label{fig:kOverLine}
\end{figure}

Changes of $k$, $P_k$ and $\epsilon_k$ can be summarised by tracking them along the streamline starting just above the upstream separating point shown in figure \ref{fig:Line}. This is visualised in figure \ref{fig:kOverLine} where both $k$ and $\xi_k$ are plotted as a function of the curvilinear coordinate $s$ normalised with the length $\ell$ of the streamline: 
\begin{equation*}
s = \frac{1}{\ell} \int_{0}^{\ell} ds \ \ \text{with} \ \ ds = \sqrt{dx^2 + dy^2};
\end{equation*}
For $ s \le 0.15$ both $\xi_k$ and $k$ are almost null, indicating the negligible turbulent activity in the first portion of the separating shear layer. For larger $s$ $\xi_k$ and $k$ show an abrupt increase due to the occurrence of the Kelvin--Helmholtz instability. The largest values of $k$ are slightly delayed compared to the largest values of $\xi_k$ due to the action of the mean convection. Then, at larger $s$, both $k$ and $\xi_k$ decrease again. In the rounded cases these variations are shifted towards larger $s$, consistently with the picture of a delayed turbulent activity. This results in a lower intensity of both $k$ and $\xi_k$ before their maxima, but also in larger intensities at larger $s$. After the reattachment point $\xi_k$ features negative values and becomes a sink for $k$. Consistently, $k$ decreases in this region and reaches its minimum in correspondence of the TE. Then, when the streamline passes over the TE, positive $\xi_k$ are observed again in correspondence of the shear layer, while negative $\xi_k$ are observed downstream, where there is no production of turbulent fluctuations. This results first in an increase of $k$, followed by a further decrease.

\subsection{Discussion}

The present results provide a picture of curvature-related effects that is not entirely in agreement with that emerging from the similar study by Rocchio et al. \cite{rocchio-etal-2020}. Indeed, the two studies possess significant differences, with the latter being based on implicit LES at the much higher $Re=40000$. The range of curvature radii considered in \cite{rocchio-etal-2020} is quite wide, spanning from $R/D=1/20$ to $R/D=1/270$, but this should be considered in view of their significantly higher $Re$. With this premise, perhaps the main finding of \cite{rocchio-etal-2020} is that even a tiny curvature radius produces an abrupt change of the mean flow, and a sudden enlargement of the primary vortex. On the contrary, our results indicate that a small amount of rounding does not affect the flow significantly, and that curvature effects manifest themselves gradually when the curvature radius increases. Moreover, we have found that the rounding affects the extension of the primary vortex in the opposite way, as in our experiments its size diminishes. In terms of turbulent kinetic energy distribution, though, there is qualitative agreement between \cite{rocchio-etal-2020} and the present results, with differences in the amount of rounding-induced changes that may simply be due to the different $Re$. 

Assessing the reason(s) for these differences certainly requires further studies. However, some hypotheses can be put forward.
As observed by \cite{lamballais-etal-2010} rounding the LE corners has two opposite effects. The downstream shift of the Kelvin--Helmholtz instability tends to increase the length of the primary vortex, while the decrease of the separation angle at the LE tends to decrease it. In the present DNS the latter effect seems to prevail, while in \cite{rocchio-etal-2020} the former one seems to dominate.
Moreover, Ref. \cite{rocchio-etal-2020} interestingly describes the appearance of a region with larger $k$ near the LE not only over the separated shear layer, but also over the front face of the body, before the separation point. To explain this observation, that cannot be attributed to an upstream shift of the shear layer instability, the authors of Ref. \cite{rocchio-etal-2020} mention that the sharp corner might introduce a large amount of $k$ which is not fully damped in their implicit LES simulation. This suggests the possibility that, at least partially, the large sensitivity to the curvature radius reported in Ref. \cite{rocchio-etal-2020} is associated to the specific numerical approach. Indeed, the proper description of a sharp corner, especially at high $Re$, requires extremely fine grids, which is the very motivation for our accounting of the geometrical singularity analytically (see next Sec. \ref{sec:correction}).
A further partial explanation of the discrepancy resides in the different numerical noise produced by the numerical methods used here and in \cite{rocchio-etal-2020}, coupled with the large receptivity of this flow to inflow perturbations. Indeed, it is well known that numerical errors and interaction between inlet and outlet boundary conditions can be artificial sources of inflow perturbations, whose level depends on the accuracy of the numerical method. The receptivity of this flow on the inflow perturbations has been largely studied to address the discrepancy between numerical simulations and experiments. For example \cite{ricci-etal-2017} via LES simulations found that a higher level of incoming turbulence corresponds to a shorter primary vortex and to an upstream shift of the secondary vortex. \cite{lamballais-etal-2010} observed that this receptivity increases with the curvatures radius and report that for $R/D=1/2$ the size of the primary vortex decreases of approximately $60\%$ compared to the case without perturbations. 

{\em Per se}, the present results are self-consistent, and compare well with other similar numerical analyses of low-$Re$ flow around bluff bodies with rounded LE. For example, \cite{lamballais-etal-2008} studied a three-dimensional D-shaped body with rounded LE at $Re=2500$, and considered two relatively large curvature radii, i.e. $R/D=1/2.5$ and $R/D=1/5$. In qualitative agreement with our results, they observed that increasing $R$ leads to a decrease of both longitudinal and vertical extensions of the main recirculating region over the longitudinal body side. \cite{lamballais-etal-2010} performed two-dimensional and three-dimensional DNS of the flow past a flat plate with rounded leading edge at $Re=4000$, with curvature radius ranging between $R/D=1/2$ and $R/D=1/16$. Their three-dimensional simulations confirm that for larger $R$ the extension of the primary vortex decreases. Moreover, a slight downstream shift of the secondary vortex is observed, which is in line with our results for the $R/D=1/64$ case, together with a decrease of the slope of the separating shear layer. They also find a decrease of the backflow in the region close to the plate side, in agreement with our simulations. Their two-dimensional simulations, instead, show completely different results, but it is known that at their high $Re$ the flow is strongly unstable to three-dimensional perturbations. 
In contrast, we have conducted preliminary two-dimensional simulations in the laminar regime at $Re=500$, and these confirm our results in the turbulent regime and agree with the three-dimensional simulations at the same Reynolds number. This is because at $Re=500$ the three-dimensionality of the flow almost does not affect the mean flow. Indeed, we have observed that the $Re=500$ only slightly exceeds the critical Reynolds number for the first onset of the first three-dimensional instability.
Finally, \cite{cimarelli-franciolini-crivellini-2020} perform LES simulations of a flat plate at $Re \approx 3000$ with both sharp and rounded LE with $R/D=1/2$. Again, their results qualitatively confirm that in the rounded configuration the extension of the main recirculating region decreases.

\section{Corner correction}
\label{sec:correction}

The two upstream LE corners where the flow impinges before separating are nominally sharp, and as such constitute a geometrical singularity that locally impacts the solution accuracy, to an extent that depends on the local fineness of the adopted grid. To overcome this, one may analytically determine the solution near the corner. In this work we follow an idea originally introduced by Luchini \cite{luchini-1991} and later taken up and expanded in \cite{burda-etal-2012}. The strategy leverages the fact that, close enough to the corners, viscous effects dominate as the velocity gradients become infinitely large. As a result, locally the in-plane velocity components (i.e. $u$ and $v$) can be deduced from the Stokes equations, where the non-linear convective terms of the Navier--Stokes equations are discarded.

\subsection{Formulation}

\begin{figure}
\centering
\begin{tikzpicture}[scale=5]	

\coordinate (O) at (0,0);
\coordinate (Ox) at (1,0);
\coordinate (Oy) at (0,-0.25);
\coordinate (P) at (-0.25,0.25);

\draw[very thick,black] (O)--(Ox);
\draw[very thick,black] (O)--(Oy);
\filldraw (-0.25,0.25) circle (0.2pt);
\node at (-0.3,0.25) {$P$};
\draw[->] (O)--(P);
\draw[->] (0.2,0) arc (0:135:0.2);
\node at (0.075,0.15) {$\theta$};
\node at (-0.125,0.08) {$r$};

\coordinate (Oo) at (-0.5,-0.25);
\coordinate (Oox) at (-0.3,-0.25);
\coordinate (Ooy) at (-0.5,-0.05);

\draw[->] (Oo)--(Oox);
\draw[->] (Oo)--(Ooy);

\node at (-0.3,-0.20) {$x$};
\node at (-0.55,-0.05) {$y$};

\end{tikzpicture}
\caption{Sketch of the polar coordinate system $(r,\theta)$ used to derive the analytical solution in the neighbourhood of the top LE corner.}
\label{fig:top_LE_corner}
\end{figure}
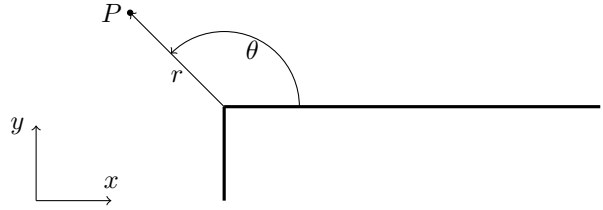

One begins by solving the Stokes equation within the portion of plane identified by two semi-infinite and perpendicular straight lines. In the following discussion we only describe the case of the top LE corner, shown in figure \ref{fig:top_LE_corner}, but the same procedure holds with obvious modifications for the bottom LE corner. The two-dimensional Stokes equation is written in a longitudinal, $z=const$ plane in terms of vorticity $\omega$ and streamfunction $\psi$, with a polar coordinate system $(r,\theta)$ such that $x=r \cos(\theta)$ and $y=r\sin(\theta)$. The equations read:
\begin{equation}
\begin{aligned}
\frac{\partial^2 \psi}{\partial r^2} + \frac{1}{r} \frac{\partial \psi}{\partial r} + \frac{1}{r^2} \frac{\partial^2 \psi}{\partial \theta^2} & = - \omega \\
\frac{\partial^2 \omega}{\partial r^2} + \frac{1}{r} \frac{\partial \omega}{\partial r} + \frac{1}{r^2} \frac{\partial^2 \omega}{\partial \theta^2} & = 0
\end{aligned}
\label{eq:Stokes_polar}
\end{equation}
where the radial and azimuthal velocity components, $u_r$ and $u_\theta$ are related to $\psi$ by:
\begin{equation}
u_r = \frac{1}{r} \frac{\partial \psi}{\partial \theta}, \ \ \ \ u_\theta = - \frac{\partial \psi}{\partial r}.
\label{eq:eq_vel_psi}
\end{equation}
We look for separable solutions in the form:
\begin{equation}
\begin{aligned}
&\psi(r,\theta)=P(r)F(\theta)
&\omega(r,\theta)=R(r)G(\theta).
\end{aligned}
\end{equation}
By introducing the functional form above into (\ref{eq:Stokes_polar}), and requiring the solutions to be regular when the corner is approached i.e. $r \rightarrow 0$, one obtains
\begin{equation}
P(r) = r^{(K+2)}; \qquad R(r) = r^K
\end{equation}
and the Stokes problem reduces to the following pair of ODEs:
\begin{equation}
G''(\theta)+K^2G(\theta)=0 \ \ \ \ F''(\theta) + (K+2)^2 F(\theta) = G(\theta).
\label{eq:diff_eq}
\end{equation}
The boundary conditions require the velocity to be zero on the two straight sides of the corner; in polar coordinates, this translates into:
\begin{equation}
\frac{\partial \psi}{\partial \theta}(r,0)=0; \quad \psi(r,0)=0; \quad \frac{\partial \psi}{\partial \theta}(r,\frac{3\pi}{2})=0; \quad \psi(r,\frac{3 \pi}{2})=0.
\label{eq:bc}
\end{equation}
Solving the two ODEs (\ref{eq:diff_eq}) leads to:
\begin{equation}
\begin{aligned}
&G(\theta)=A_1 \cos\left( K \theta \right) + A_2 \sin \left( K \theta \right) \\ 
&F(\theta) = B_1 \cos \left( (K+1) \theta \right) + B_2 \sin \left( (K+2) \theta \right) + B_3 \cos \left(K\theta \right) + B_4 \sin \left( K \theta \right) .
\end{aligned}
\end{equation}
where $A_1$, $A_2$, $B_1$, $B_2$, $B_3$, $B_4$ are constants to be determined via the boundary conditions \ref{eq:bc}. A linear system is obtained:
\begin{equation*}
M(\gamma)\vect{b}=\vect{0}
\end{equation*}
where $\gamma=K+1$, and $\vect{b}$ is the vector of the unknowns $\vect{b}=(B_1,B_2,B_3,B_4)$ needed to determine $\psi$ and therefore $u_r$ and $u_\theta$. To solve the system we require that
\begin{equation*}
det \left( M(\gamma) \right) =0
\end{equation*}
which leads to the following relation:
\begin{equation}
\gamma^2 - \sin^2\left(\gamma \frac{3\pi}{2} \right)=0.
\end{equation}
The numerical solution of this equation via bisection yields $\gamma \approx 0.5444837$. Then by solving the linear system, and by taking $B=1$ without loss of generality thanks to the linearity of the problem, the unknowns $\vect{b}$ are obtained:
\begin{equation}
\begin{aligned}
B_1 & = 1 \\
B_2 & = \frac{ \left( \gamma-1 \right) \cos \left( (\gamma-1) \frac{3\pi}{2} \right) - (\gamma-1) \cos \left( (\gamma-1) \frac{3\pi}{2} \right) }
             { \left( \gamma-1 \right) \sin \left( (\gamma+1) \frac{3\pi}{2} \right) - (\gamma+1) \sin \left( (\gamma-1) \frac{3\pi}{2} \right) } = D_2(\gamma) \\
B_3 & = - 1\\
B_4 & = \frac{ \left( \gamma+1 \right) \cos \left( (\gamma+1) \frac{3\pi}{2} \right) - (\gamma+1) \cos \left( (\gamma-1) \frac{3\pi}{2} \right) }
             { \left( \gamma-1 \right) \sin \left( (\gamma+1) \frac{3\pi}{2} \right) - (\gamma+1) \sin \left( (\gamma-1) \frac{3\pi}{2} \right) } = D_4(\gamma) .
\end{aligned}
\end{equation}

Once the asymptotic behaviour of $\psi(r,\theta)$ in the vicinity of the corner has been determined, we get $u_r(r,\theta)$ and $u_{\theta}(r,\theta)$ by their definitions (\ref{eq:eq_vel_psi}):
\begin{equation}
\begin{aligned}
u_r(r,\theta) = r^\gamma \large( (\gamma-1) \sin \left( (\gamma-1) \theta \right) -
                                      (\gamma+1) \sin \left( (\gamma+1) \theta \right) & + \\
                                       D_4(\gamma) (\gamma-1) \cos \left( (\gamma-1) \theta \right) +
                                      D_2(\gamma) (\gamma+1) \cos \left( (\gamma+1) \theta \right) & \large),
\end{aligned}
\end{equation}
\begin{equation}
\begin{aligned}
u_{\theta}(t,\theta) = - (\gamma+1) r^\gamma \large( & \cos \left( (\gamma+1) \theta \right) +
                                                        D_2(\gamma) \sin \left( (\gamma+1) \theta \right) - \\
                                                      &  \cos \left( (\gamma-1) \theta \right) +
                                                        D_4(\gamma) \sin \left( (\gamma-1) \theta \right) \large).
\end{aligned}
\end{equation}
Last, the Cartesian velocity components $u$ and $v$ can be easily retrieved by:
\begin{equation}
u = u_r \cos(\theta) - u_{\theta} \sin(\theta); \qquad v = u_r \sin(\theta) + u_{\theta} \cos(\theta). 
\end{equation}

Pressure is obtained from the Stokes equation in polar coordinates solved for $\partial p / \partial r$, i.e.:
\begin{equation}
\frac{\partial p}{\partial r} = \nu \left( \frac{\partial^2 u_r}{\partial r^2} + \frac{1}{r} \frac{\partial u_r}{\partial r} 
                                         + \frac{1}{r^2} \frac{\partial^2 u_r}{\partial \theta^2} 
                                         - \frac{2}{r^2}\frac{\partial u_{\theta}}{\partial \theta} - \frac{u_r}{r^2} \right).
\end{equation}
An integration in $r$ yields:
\begin{equation}
p(r,\theta) = 4 \nu \gamma r^{\gamma-1} \left (D_4(\gamma) \cos \left( (\gamma-1) \theta \right) - D_3(\gamma) \sin \left( (\gamma-1) \theta \right) \right).
\end{equation}

Once the correct local behaviour of $u$, $v$ and $p$ in the vicinity of the corner is  analytically determined, this information is used in the DNS code, in such a way that the DNS solution possesses the required characteristics. The general idea is to use the exact Stokes solution to enforce a deferred correction of the DNS solution via  two correction terms, one for the momentum equations in the $x$ direction and one for that in the $y$ direction, that are used at each iteration to ensure that the updated solution satisfies the Stokes equation in the vicinity of the corner. In the following the region near the corner interested by the correction is defined by:
\begin{equation*}
(x-x_c)^2+(x-y_c)^2 \le (0.1D)^2
\end{equation*}
where $(x_c,y_c)$ are the coordinates of the corner. In the present work we choose to apply the correction within a distance of $0.1 D$ from the corner, that is enough to let the correction decrease to zero, but it must be noted that the distance needs to be accurately tuned, to avoid an incomplete correction due to an excessively short distance.

\subsection{Results}

\begin{figure}
\centering
\includegraphics[trim=20 0 200 400,clip,width=1\textwidth]{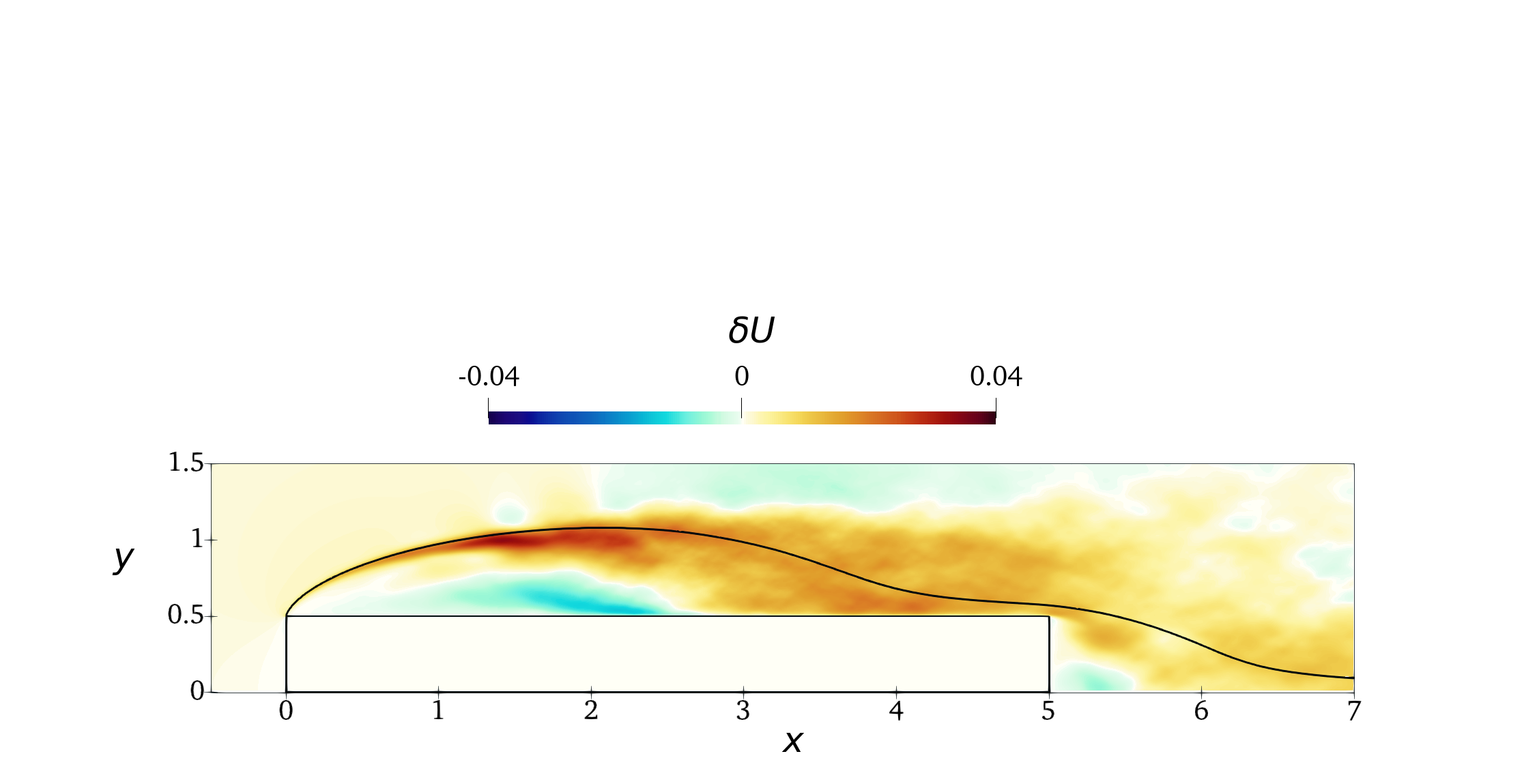}
\includegraphics[trim=20 0 200 400,clip,width=1\textwidth]{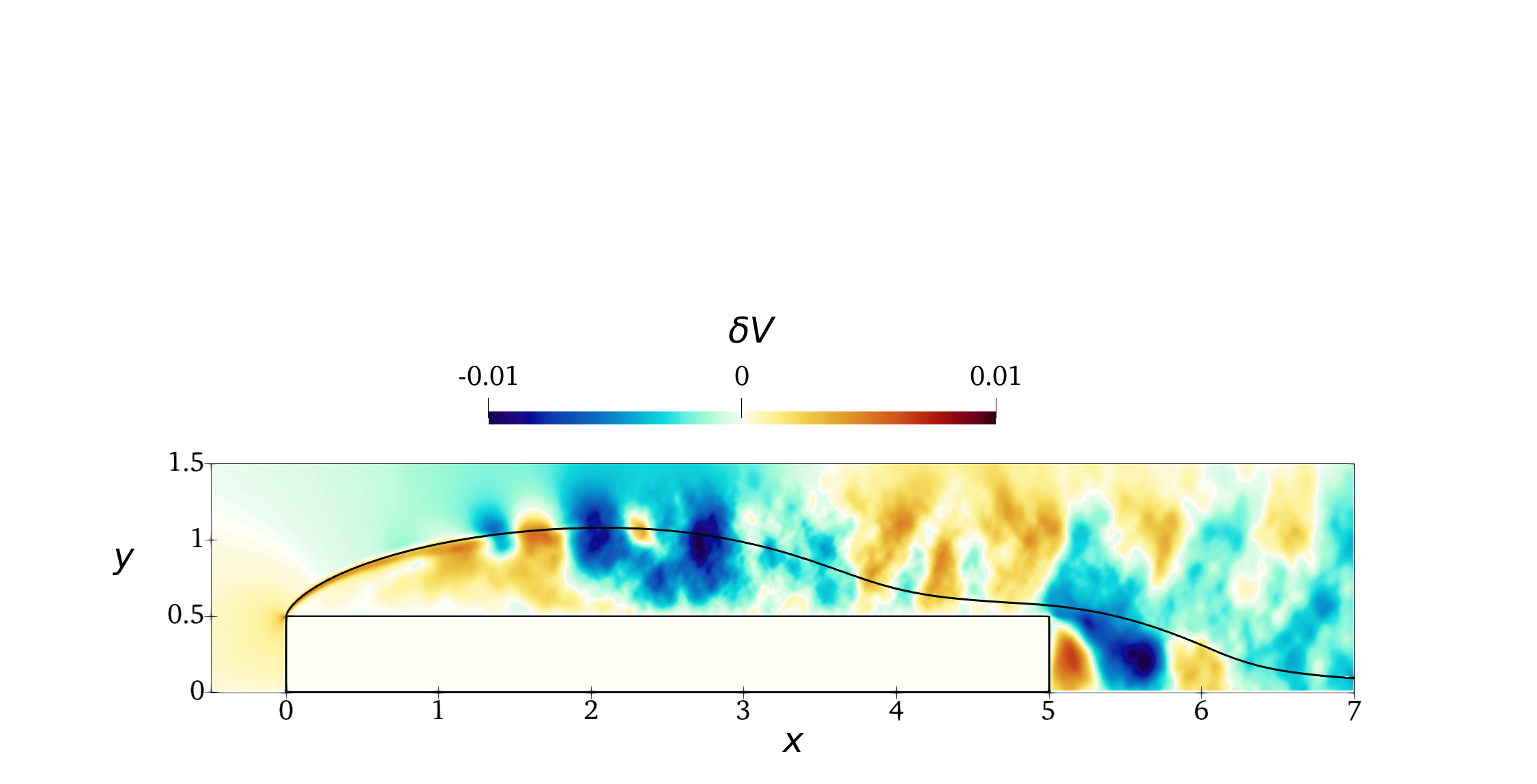}
\includegraphics[trim=0 0 1300 600,clip,width=0.46\textwidth]{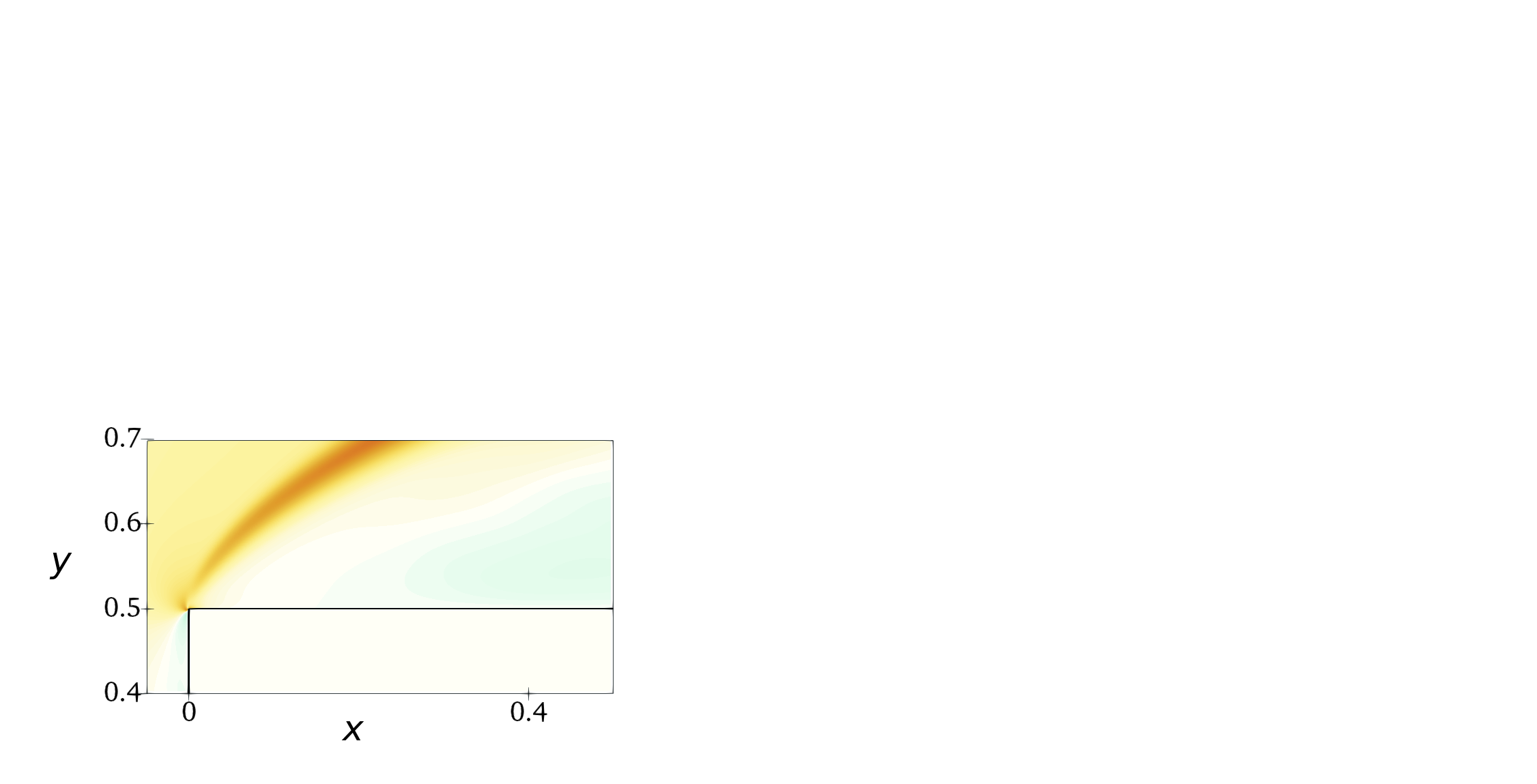}
\includegraphics[trim=0 0 1300 600,clip,width=0.46\textwidth]{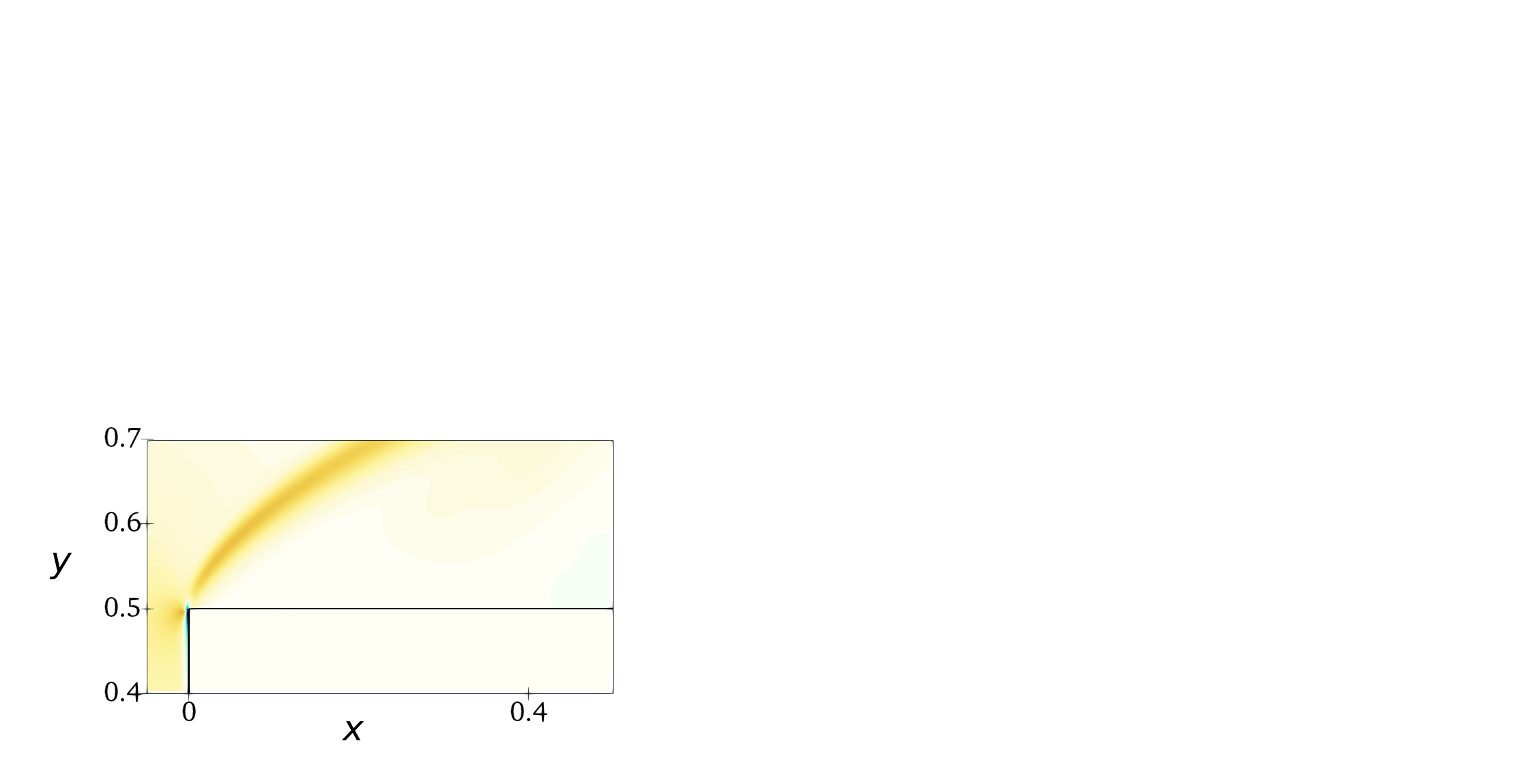}
\caption{Effect of the analytical correction of the LE corner singularity on the mean flow. Top panel is for $\delta U=U_c-U_r$ while the central one for $\delta V = V_c-V_r$, where subscripts $\cdot_c$ and $\cdot_r$ refer to the corrected and the reference case. The black line, drawn for the reference case, indicates the streamline originating from just above the LE corner and locates the shear layer. The bottom panels show a zoom on the LE corner for $\delta U$ (left) and $\delta V$ (right).}
\label{fig:deltaVelocity}
\end{figure}

The improvement made possible by the analytical correction for the LE corner singularity are considered here in terms of mean flow. Since the nominal geometry is the same, they can be simply expressed in terms of the difference between the two mean fields, with and without correction. Note that the very fine grid of the reference flow is used, and this precludes observing potentially large benefits. The goal of the present work is only the qualitative description of the effects induced by the analytical correction. 

Figure \ref{fig:deltaVelocity} plots the difference of the mean velocity field in terms of its two components,
\begin{equation*}
\delta U = U_{c} - U_{r} \ \text{and} \ \delta V = V_c-V_r,
\end{equation*}
where the subscripts $\cdot_c$ and $\cdot_r$ refer to the case with the analytical correction and to the reference case without it. The following discussion only considers the top side of the cylinder, but the same observations are valid for the bottom side too, by suitably accounting for the symmetries of the flow. The black solid line is the streamline of the reference case starting just above the top LE corner at $(x,y) = (0,0.5001)$, and is useful to locate the shear layer. The maps of $\delta U$ and, to a larger extent, $\delta V$ both reveal the presence of a small amount of statistical noise, due to the finite temporal average. However, the robustness of the observations has been checked by computing the maps with only half of the sample size; the noise is correspondingly increased, but the qualitative scenario remains unchanged.

\begin{figure}
\centering
\includegraphics[trim=0 0 1300 600,clip,width=0.6\textwidth]{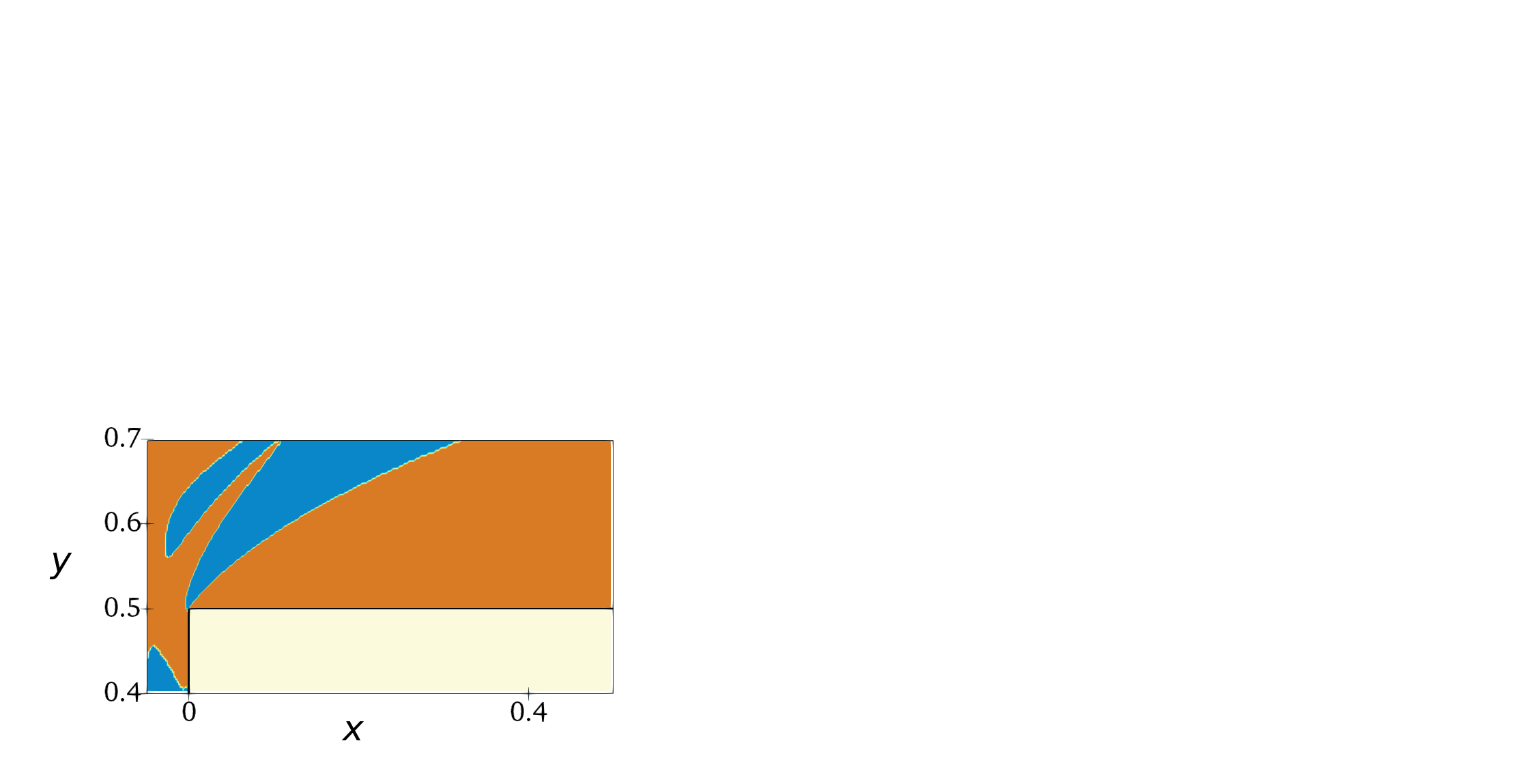}
\caption{Sign of the change $\delta (\text{d}y_s/\text{d}x)$ in the local slope of mean streamlines. Blue: negative change; orange: positive change.}
\label{fig:diff_slope}
\end{figure}

Near the LE corner, where the analytical correction is directly applied, both $\delta U$ and $\delta V$ are positive in the shear layer. On the other hand, just before the corner there are two tiny regions with $\delta U<0$ and $\delta V<0$ attached to the vertical side of the cylinder. This indicates that the corrected flow is decelerated just before its impingement on the corner, whereas the shear layer separating from it is accelerated. The maps of $\delta U$ and $\delta V$ show that in the flow with correction the streamlines more closely follow the geometry of the corner, as explained in the following discussion. The local slope of a mean streamline $y_s(x)$ is defined as:
\begin{equation*}
\frac{\text{d}y_s}{\text{d}x} = \frac{V}{U}.
\end{equation*} 
Therefore the slope change induced by the correction is
\begin{equation*}
\delta \left( \frac{\text{d}y_s}{\text{d}x} \right) = 
\frac{V_c}{U_c} - \frac{V_r}{U_r} = 
\frac{U_r V_c - U_c V_r}{U_c U_r}.
\end{equation*}
This quantity is plotted in figure \ref{fig:diff_slope}, where blue indicates $\delta (\text{d}y_s/\text{d}x)<0$ and orange $\delta (\text{d}y_s/\text{d}x)>0$.
Just before the corner $\delta (\text{d}y_s/\text{d}x)$ is positive for $0.4 \le y \le 0.5$; note that this vertical extension almost corresponds to the distance chosen for the correction to apply. This indicates that the corrected streamlines are more vertical in this region and therefore more aligned with the vertical side of the cylinder. On the other hand, the blue region after the corner indicates that the streamlines, after crossing the LE, have a lower slope and align better to the longitudinal side of the cylinder. In other words, the separation angle of the shear layer decreases once the analytical correction is used.

The analytical correction acts locally near the corners, but its effects are seen also further from it, in the entire region above the cylinder side. For example, similarly to what previously observed for the rounded configurations, the decrease of the separation angle of the shear layer results in a slight decrease of both the vertical and longitudinal extent of the primary vortex, with $L_1$ dropping from $3.955$ to $3.92$ ($\approx -1.3 \%$). The top panel of figure \ref{fig:deltaVelocity} indicates that within the primary vortex the analytical correction yields $\delta U>0$ almost everywhere but close to the cylinder side for $x \ge 2.5$, where $\delta U <0$ shows an increase of the backflow. The positive $\delta U>0$ in the region around the limiting streamline is consistent with a smaller vertical extent of the primary vortex. On the other hand, after the first portion of the shear layer where $\delta V > 0$, for intermediate $x$ $\delta V$ becomes negative and then positive again after the reattachment point. This once again agrees with the picture of a shorter primary vortex. Indeed this change of the sign of $\delta V$ is due to an upstream shift of the point at which the streamlines turn towards lower $y$.

The small differences between the reference case and the case with the analytical correction confirm that the resolution used in this work near the LE corners is more than adequate. We expect that increasing the resolution further would lead eventually to vanishing differences.

\section{Conclusions}

The present work has studied via Direct Numerical Simulations (DNS) the BARC benchmark flow in the turbulent regime at $Re=3000$, with focus on the geometrical characterisation of the leading-edge (LE) corners. In doing so we intend to contribute to the discussion whether the geometrical details of the nominally sharp LE corners could explain at least partially the scatter of available data.

In the first part of the work, the effect of rounded LE corners has been studied. Two values for the curvature radius $r$ have been considered, namely $R/D=1/128$ and $R/D=64$, which mimic the unavoidable imperfections that would be present in a physical model owing to manufacturing procedures. The present investigation follows a similar one by \cite{rocchio-etal-2020}, who simulated the flow via Large Eddy Simulations at a much larger Reynolds number, i.e. $Re=40000$. We have discussed the possible reasons for differences between that study and the present results, which may be traced down to the significantly different Reynolds number combined with the different modelling approach and numerical method. Unlike in \cite{rocchio-etal-2020}, we have found that a small amount of rounding does not abruptly change the features of the mean flow, and that the effects increase gradually with $r$. In the rounded configurations, the shear layer separates from the LE with a milder slope, so that the vertical and longitudinal sizes of the main recirculating region are both reduced. Interestingly, rounding the LE corners has been found to affect the wake vortex too, with its longitudinal size slightly increasing with $R$. This is explained by the observation that rounding the LE increase  the velocity over the last part of the cylinder side after the reattachment point, resulting in a faster TE shear layer. 

The inspection of the turbulent kinetic energy reveals that in the rounded cases the turbulent activity is slightly delayed in the downstream direction compared to the sharp configuration, so that $k$ is decreased in the first part of the cylinder, but slightly increased in the second part. This happens via the downstream shift of the Kelvin--Helmholtz instability of the shear layer, accompanied by a similar shift of the transport mechanism involving the upstream portion of the primary vortex. A partial explanation of the slight increase of $k$ in the second part of the cylinder side resides in the larger increase of the production $P_k$ of turbulent kinetic energy compared to its dissipation rate $\epsilon_k$, resulting in an overall increase of the source term $\xi_k = P_k - \epsilon_k$. Once again, the present results do not fully agree with findings reported by \cite{rocchio-etal-2020}. They observed a spatially delayed development of the turbulent activity too, but in their simulation this produces a large increase of the longitudinal size of the primary vortex already for very small $R$.

The second part of the work restores the LE corners to their nominal sharp geometry. For the first time in the BARC context, we use an analytical solution of the Stokes flow over a sharp corner (see \cite{moffat-1964,luchini-1991}) to locally improve the accuracy of the DNS numerical solution, which is unavoidably degraded by the geometrical singularity. We have outlined a strategy based on the idea that, in the vicinity of the corner, the in-plane velocity components must obey the Stokes equations, as viscous effects are dominant. By applying a numerical correction to the DNS solution such that near the LE corners the Stokes solution is recovered, we have described how the mean flow appears to better adapt to the corner shape and becomes more aligned to the cylinder sides. As a result, the shear layer separates from the LE with a milder angle, therefore yielding a decrease of the size of the primary vortex. It should be noted that, in the present work, the analytical correction has been applied to a well-resolved DNS only. As a consequence, the improvements are small in magnitude. Further work is needed to properly characterize and assess the performance of the method. However, the true value of the approach can be appreciated when the correction is enforced to coarser-grid simulations, where it should allow a significantly lower computational cost for a given accuracy. 

\section*{Acknowledgments}
Computing time has been provided by the Italian supercomputing center CINECA under the ISCRA C projects TAWBF and AGKEbump. 

\section*{Conflict of interest}
The authors declare that they have no conflict of interest.

\bibliographystyle{plain}

\end{document}